\documentclass[useAMS,usenatbib,fleqn]{mnras}
\usepackage{times}
\usepackage{graphicx}
\usepackage{epstopdf}
\usepackage{amsmath}
\usepackage{amsfonts}
\usepackage{amssymb}
\usepackage{hyperref}
\usepackage[usenames, dvipsnames]{color}
\usepackage{lipsum}
\usepackage{pdflscape}
\usepackage{xcolor}
\usepackage[normalem]{ulem}

\newcommand{\Mi}{M_{\rm i}}
\newcommand{\rh}{r_{\rm h}}
\newcommand{\rj}{r_{\rm J}}
\newcommand{\kpc}{{\rm kpc}}
\newcommand{\gyr}{{\rm Gyr}}
\newcommand{\rg}{R_{\rm G}}

\newcommand{\feh}{\[Fe/H\]}
\newcommand{\rapo}{R_{\rm apo}}
\newcommand{\rperi}{R_{\rm peri}}
\newcommand{\mud}{\mu_{\rm dyn}}
\newcommand{\msun}{{\rm M}_\odot}
\newcommand{\kms}{{\rm km}\,{\rm s}^{-1}}

\def\deg{^{\circ}}
\newcommand{\comm}[1]{#1}

\def\fifm{f^{\rm ifm}}
\def\feh{{\rm [Fe/H]}}
\def\md{m_{\rm d}}
\def\sigmat{\sigma_t}
\def\tmsright{\tau_{\rm ms,r}}
\def\tmsleft{\tau_{\rm ms,l}}
\newcommand{\mi}{m_{\rm i}}
\newcommand{\mj}{m_j}

\newcommand{\dr}{{\rm d}}
\newcommand{\emacss}{{\sc emacss}}
\newcommand{\nbin}{n_{\rm bin}}
\newcommand{\nj}{N_j}
\newcommand{\njs}{N_j^{\rm s}}
\newcommand{\njr}{N_j^{\rm r}}
\newcommand{\njdot}{\dot{N}_j}
\newcommand{\njrdot}{\dot{N}_j^{\rm r}}
\newcommand{\njsdot}{\dot{N}_j^{\rm s}}
\newcommand{\njescdot}{\dot{N}_j^{\rm  esc}}
\newcommand{\njsescdot}{\dot{N}_j^{\rm s, esc}}
\newcommand{\njrescdot}{\dot{N}_j^{\rm r, esc}}
\newcommand{\njssevdot}{\dot{N}_j^{\rm sev}}
\newcommand{\nkrdot}{\dot{N}_k^{\rm r}}

\newcommand{\tms}{\tau_{\rm ms}}
\newcommand{\mto}{m_{\rm TO}}

\title[The devil is in the tails]{The devil is in the tails: the role of
globular cluster mass evolution on stream properties}
\author[Balbinot \& Gieles]{Eduardo Balbinot\thanks{e.balbinot@surrey.ac.uk} 
                            and Mark Gieles \\
                            Department of Physics, University of Surrey, Guildford GU2 7XH, UK
\\}

\begin{document}
\maketitle
\label{firstpage}
\pagerange{\pageref{firstpage}--\pageref{lastpage}}

\begin{abstract}
    We present a study of the effects of collisional dynamics on the formation
    and detectability of cold tidal streams. A semi-analytical model for the
    evolution of the stellar mass function was implemented and coupled to a fast
    stellar stream simulation code, as well as the synthetic cluster evolution
    code {\sc emacss} for the mass evolution as a function of a globular cluster
    orbit. We find that the increase in the average mass of the escaping stars
    for clusters close to dissolution has a major effect on the observable
    stream surface density. As an example, we show that Palomar 5 would have
    undetectable  streams (in an SDSS-like survey) if it was currently three
    times more massive, despite the fact that a more massive cluster loses stars
    at a higher rate. This bias due to the preferential escape of low-mass stars
    is \comm{an alternative} explanation for the absence of tails near massive
    clusters, than a dark matter halo associated with the cluster.  We explore
    the orbits of a large sample of Milky Way globular clusters and derive their
    initial masses and remaining mass fraction. Using properties of known tidal
    tails we explore regions of parameter space that favour the detectability of
    a stream. A list of high probability candidates is discussed.
\end{abstract}


\section{Introduction}

Stellar streams are the long promised probes of the gravitational potential in
galaxies. In the Milky Way (MW) these streams are visible as resolved stellar
populations and upon the advent of large photometric surveys such as the Sloan
Digital Sky Survey \citep[SDSS;][]{sdss} and the Dark Energy Survey
\citep[DES;][]{des} began to be discovered in large numbers. These surveys
revealed two types of streams: hot streams from disrupted high velocity
dispersion, dark matter dominated dwarf galaxies, and cold streams from existing
or completely dissolved globular clusters (GCs). The most remarkable examples of
each type are the Sagittarius stream, a hot stream covering most of the sky
\citep{Newberg02} and the  tidal tails of Palomar 5, a 22$\deg$ long cold
stellar stream \citep{Odenkirchen01, Grillmair06, Bernard16}.

While both kinds of streams are promising proxies of the gravitational potential
of the MW \citep[e.g.][]{Koposov10, 2014MNRAS.445.3788G}, cold streams are
appealing because of their simplicity and long phase mixing time
\citep{Helmi99}. The progenitors of cold streams are GCs which have internal
velocity dispersion ($\lesssim 10\,\kms$), much lower than their orbital
velocity (a few $100\,\kms$). Stars that become unbound follow approximately the
same orbit \comm{(with some offset, as shown by \citealt{Eyre09})}. Each unbound
star can be seen as a test particle for the MW gravitational potential while the
collection of escaper stars should approximately trace a single orbit in the
underlying potential \citep{Koposov10}. 

Several studies fail to find convincing signs of tails around MW GCs
\citep{Leon00, Kuzma16}. The absence of tails may point at the presence of a
dark matter halo surrounding the GC, preventing the stars from escaping
\citep{1996ApJ...461L..13M}.  The lack of tails associated with a GC could also
be because the GC is in a weak tidal field. This may explain the absence of
tails from GCs at large Galactic centre, but not why there are no cold streams
associated with massive GCs. From theory and numerical results we know that
\comm{the mass-loss rate as the result of evaporation depends on the cluster mass
$M$ and the Galactocentric distance $\rg$ as}: $\dot{M} \propto -M^{1/4}/R_{\rm
G}$, for clusters evolving in an isothermal halo \citep*{Baumgardt01, Gieles11}.
However, the most prominent cold streams observed are either progenitor-less or
from low-mass GCs. One possible observational bias, proposed by \citet{Leon00},
is the preferential loss of low-mass stars, which makes streams fainter. Indeed,
observations show many GCs have mass functions that are depleted in low-mass
stars \citep[see e.g.][]{Paust10}.

In this paper we further investigate the interplay between the internal dynamics
of GCs and its effect on the formation and structure of tidal tails. We focus on
the preferential loss of low-mass stars due to mass segregation in the cluster
and the evolution of the velocity dispersion of potential escaper stars,
\comm{i.e. stars that are energetically unbound, but still association with the
cluster \citep[see][]{2000MNRAS.318..753F}.}

This paper is organized as follows: in section 2 we describe a method to evolve
a GC mass function as it loses mass while orbiting the Galaxy. In section 3 we
present a simple model for the mass evolution of a cluster in the Galaxy as well
as up-to-date orbital parameters and mass estimates for a large sample of MW
GCs. In section 4 we explore the conditions for the observability of cold
streams and highlight noteworthy candidates for the search of streams. In
section 5 we use a spray-particle tidal tail simulation of Palomar 5 to
illustrate the impact of the preferential loss of low-mass stars in the
visibility of tidal tails. In section 6 we give our conclusions in the light of
upcoming surveys. 

\section{Method}
\label{sec:evolve_mf}
Ideally, we would like to map progenitor properties (i.e. initial mass and
orbit) to stream properties such as length, width, particle mass, and velocity
dispersion. In the case of GCs this comes at the cost of $N$-body models which
are computationally expensive. To avoid the complications of such models, we
choose to use the fast cluster evolution code Evolve Me A Cluster of
StarS\footnote{Available from
\href{https://github.com/emacss/emacss}{https://github.com/emacss/emacss}}
\citep[\textsc{emacss},][]{emacsscode, 2014MNRAS.437..916G, Alexander14}
\comm{\emacss\ evolves several global cluster properties, such as the total
bound mass $M$ and number of stars $N$, half-mass radius $\rh$, and parameters
describing the degree of mass segregation and the profile of the cluster via
coupled ordinary differential equations (ODEs). These ODEs include the effect of
escape of stars over the tidal boundary (evaporation), mass-loss as the result
of stellar evolution and the diffusion of energy as the result of two-body
relaxation on $M$, $N$ and $\rh$. After a number of relaxation times have
elapsed, the cluster is assumed to reach a stage of balanced evolution, in which
the rate of change of the total cluster energy $E$ is set by the flow of energy
through $\rh$, following \citet{H61}. For full details on the implementation we
refer to \citet{Alexander14}.} The code needs only a fraction of a second to
evolve a cluster until dissolution and the computational effort is
$N$-independent. The code considers the effects of mass loss and expansion of
the cluster as the result of stellar evolution, two-body relaxation and escape
of stars due to a (static) tidal field.  On top of the existing \textsc{emacss}
implementation, we build a semi-analytical prescription of the evolution of the
mass function which is described in detail in the following section.

\subsection{Mass function evolution}
In this section we present an algorithm to numerically evolve the stellar mass
function (MF). Various prescriptions for the time-dependent MF exist
\citep[e.g.][]{2009A&A...507.1409K,Lamers13}. We aim to develop a
prescription that can be coupled to the fast cluster code \emacss. Because
\emacss\ solves a set of coupled ODEs, we develop a prescription for the rate of
change of the number of stars and stellar remnants  as a function of stellar
mass $m$. We aim to correctly describe the preferential ejection of low-mass
stars as the result of two-body relaxation, and the formation of stellar
remnants as the result of stellar evolution. Below we describe the algorithm.

To evolve the MF of  stars and remnants, we define $\nbin\simeq100$
logarithmically spaced mass bins between $0.1\,\msun$ and $100\,\msun$. At $t=0$
the number of stars in each bin, $\nj$, is found from the stellar initial mass
function (IMF, $\dr N/\dr\mi$, \comm{where $\mi$ is the initial mass of a star}),
the total number of stars in the cluster $N$ and the width of the individual
bins $\Delta \mj$ as

\begin{equation}
\nj= N \left.\frac{\dr N}{\dr\mi}\right|_{m_j} \Delta \mj
\end{equation}
such that $\sum_{j=1}^{\nbin} \nj = N$ and $\sum_{j=1}^{\nbin} \nj\mj = M_{\rm
i}$, where $M_{\rm i}$ is the initial mass of the cluster. We then consider that
$\nj = \njs  + \njr$, with $\njs$ the number of stars and $\njr$ the number of
stellar remnants as a function of mass. At $t=0$ there are no remnants. 

We then solve for $\njs$ and $\njr$ by integrating expressions for the rate
of change of $\njdot = \njsdot + \njrdot$. Both stellar evolution and escape of
stars contribute to these rates, which we discuss in the next sections.

\subsection{Stellar evolution}
\label{ssec:sev}
To  evolve stars by stellar evolution, we move stars that reach the end their
main sequence life from $\njs$ to $\njr$.  We approximate  the main-sequence
lifetimes by

\begin{equation}
\tms(\mi) = 0.21\exp\left(10.4\mi^{-0.322}\right),
\label{eq:tms}
\end{equation}
which describes the main sequence life times of stars $0.5<\mi/\msun<30\,\msun$
and $\feh=-0.5$ to within 10\% compared to the SSE (Single-Star Evolution)
models of \citet{2000MNRAS.315..543H}, \comm{which provides a set of analytic
    relations that approximate the evolution of stars of different masses and
chemical composition}. \comm{In our MF evolution method,} stars are assumed to
have a constant mass $\mi$ until $\tms(\mi)$ is reached and then their mass is
reduced by a factor

\begin{align}
\fifm=
\begin{cases}
0 & \mi>10\,\msun,\\
0.56(\mi/\msun)^{-0.56} & \mi\le10\,\msun.
\end{cases}
\end{align}

Removing all stars with $\mi>10\,\msun$ corresponds to 0\% retention of black
holes. A single relation for the masses of neutron stars and white dwarfs
reproduces the results from SSE to within 10\%. The most massive neutron stars
is $1.54\,\msun$ and a solar mass star results in a $0.56\,\msun$ white dwarf.

\comm{We introduce stellar evolution by removing stars are at a rate}
\begin{align}
\njssevdot  = 
\begin{cases}
    -\njs\left(\dfrac{t-\tmsright}{\sigmat}\right), &t>\tmsright,\\
0,& t<\tmsright,
\end{cases}
\end{align}
where $\tmsright$ is the main sequence life time of a star with $m_j + 0.5\Delta
m_j$ (i.e. corresponding to the right side of  bin $j$) and $\sigmat$ sets the
speed with which the bin is emptied. From experiments we find smooth evolution
of the mass function when using $\sigmat = 0.7(\tmsright-\tmsleft)$, where
$\tmsleft = \tms(m_j-0.5\Delta m_j)$ i.e. the life time of a star with mass
corresponding to the left side of bin  $j$.

While stars are being removed, remnants are being created, hence we fill bins of
the remnants mass function at a rate

\begin{align}
\nkrdot(\mto(t)\fifm)&=- \njssevdot,
\end{align}
where $\mto(t)$ is the turn-off mass at time $t$, which is found form the
inverse of the main sequence time relation (equation~\ref{eq:tms}). The minus
sign ensures that $\nkrdot\ge0$ (note that $\njssevdot<0$). 

\subsection{Escape}
\label{ssec:escape}
\begin{figure}
\centering
\includegraphics[width=0.45\textwidth]{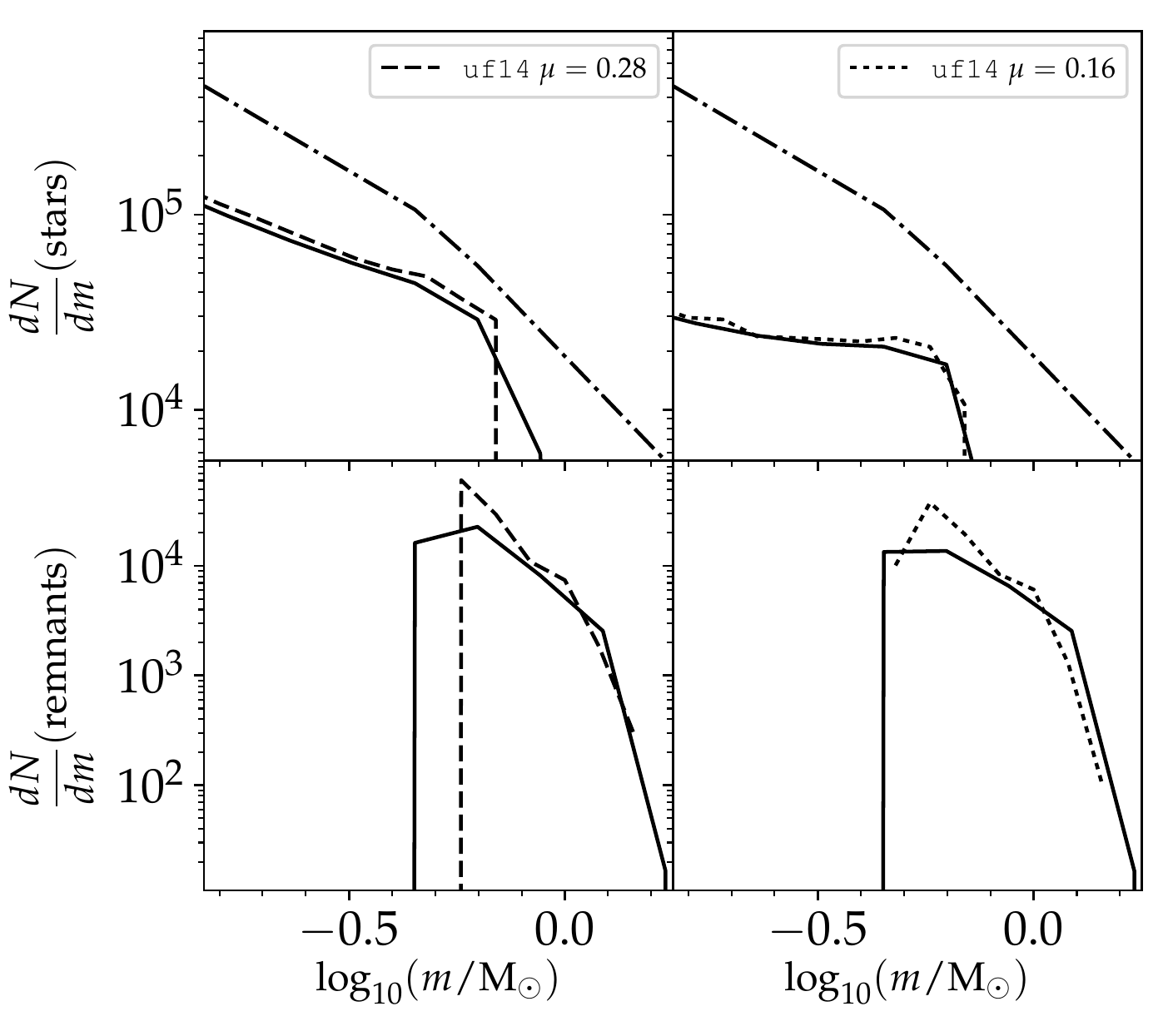}
    \caption{Comparison of the synthetic MF (solid lines) with the \texttt{uf14}
    model from \citet{Lamers13}. The top two panels show the visible stars at
    $\mu=0.28$ (left) and $\mu=0.16$ (right), \comm{where $\mu$ is the remaining
    mass fraction in the cluster.} The dot-dashed line in both top panels is the
    IMF. The two bottom panels show the MF of the stellar remnants, compared at
    the same $\mu$ as before.}
\label{fig1}
\end{figure}

In the pre-collapse phase we assume that the escape rate is independent of
stellar mass, i.e. $\njsdot = \njs \dot{N}/N$ (similarly for $\njrdot$), where
$\dot{N}$ is the total escape rate of the cluster which comes from {\sc emacss}.
After core collapse (also given by {\sc emacss}) we apply an escape rate to each
bin

\begin{align}
\njescdot  = \dot{N} \frac{\njs + \njr}{N} \frac{g(m_j)}{\sum_j g(m_j)}
\end{align}
where

\begin{align}
g(m_j)  = 
\begin{cases}
1  - \left(m_j/\md\right)^{1/2} , &m_j<\md,\\
0,                              &m_j\ge \md.
\end{cases}
\end{align}
This simple functional form for $\njescdot$ was found by matching the time
derivatives of the functional forms for the MF as a function of time given by
\cite{Lamers13}. With a value of $\md\simeq1.1\,\msun$ we find a good
agreement with results from $N$-body models (see example in Fig.~\ref{fig1}).
The 100\% retention of stars and remnants more massive than $\md$ is 
becoming problematic when less than a few \% of the initial mass is remaining,
and hence this approximation is good enough for our purpose. The escape rate is
divided over the stars and remnants as $\njsescdot = \njescdot \njs/\nj$
(similarly for $\njrescdot$), such that $\njescdot = \njsescdot + \njrescdot$.

\begin{figure}
\centering
\includegraphics[width=0.45\textwidth]{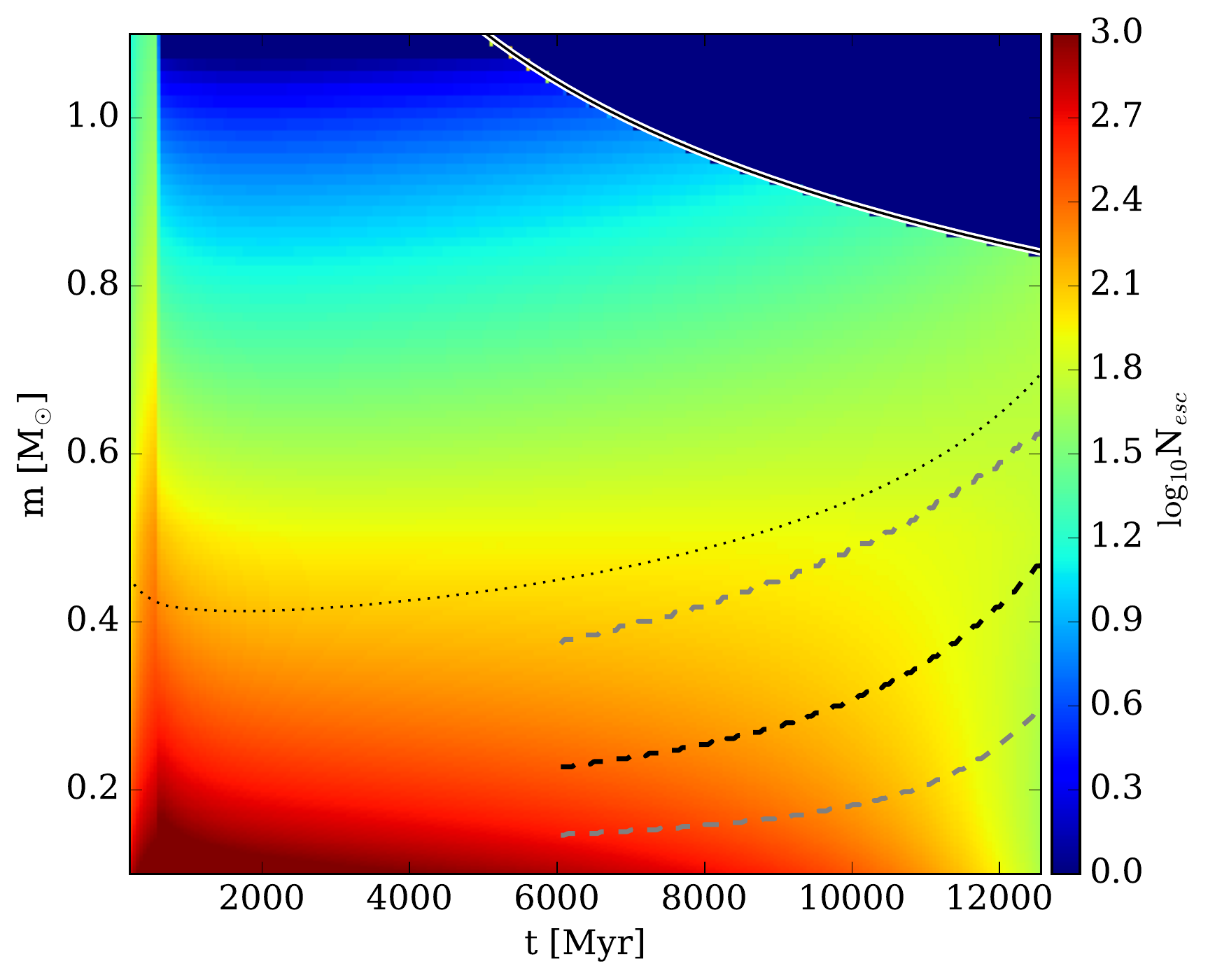}
    \caption{Number of escapers (colour coded in log-scale) as a function of
    time and mass for a simulated cluster with an initial mass of $2\times10^5$
    M$_{\odot}$ on a circular orbit at a Galactocentric distance of 18 kpc
    (similar to Palomar 5). The solid line shows the main-sequence turn-off mass
    evolution, the dark dotted line is the median stellar mass remaining in the
    cluster and the dark dashed line is the median mass of the escaper stars
    while the grey dashed lines are mark the position of the first and third
    quartiles (i.e. 50\% of the stars are between the grey lines).}
\label{fig_heat}
\end{figure}

\subsection{Validation}

To validate our implementation of the MF evolution we compare the predicted MF
with one of the $N$-body simulations of \citet{Lamers13}. These simulations are
setup with a Kroupa IMF \citep{Kroupa01} and include stellar evolution. We find
that our model provides an accurate description of the MF evolution in these
simulations.  In Fig.~\ref{fig1} we show a comparison of an $N$-body model with
$N=131\,072$ particles on a circular orbit at a radius of 8.5 kpc in a
logarithmic potential.  For illustrative purposes we compare our prediction with
the $N$-body model at two different stages of dissolution $\mu = [0.16, 0.28]$,
where $\mu$ is the remaining mass fraction of the cluster. We also show the
predicted number of \comm{stellar remnants (mainly white dwarfs)} compared to
the $N$-body results (bottom panels). We observe that the visible stars are very
well represented by our synthetic model with only minor deviations at the
turn-off mass due to our approximation of the main-sequence lifetime. The
remnant MF is in good general agreement considering the simplicity of our
method.

Fig.~\ref{fig_heat} shows a summary of the escaper MF (EMF) evolution for a
cluster that is near complete dissolution. We chose the initial conditions to
reproduce the orbit of Palomar 5 \citep{Kuepper15} and the initial mass such
that full dissolution is at $\sim$ 13 Gyr. Note that at $\sim$500 Myr there is
an abrupt change in the EMF, this happens at core-collapse where the
preferential loss of low-mass stars starts to take place. Also, as the cluster
approaches 13 Gyr, the median mass of escapers rises significantly, accounting
to more than 50\% of the stars above 0.45 $\msun$ and 25\% above 0.6 $\msun$.
This has important consequences for the visibility of streams, which we
discuss in section~\ref{sec:visibility}.

\section{Mass evolution in the Galaxy}

The framework developed here allows us to predict the mass spectrum of escaped
stars of a cluster given $\mu$. Despite being a simple parametrisation,
describing a cluster in terms of $\mu$ requires prior knowledge of its current
mass ($M$), initial mass ($\Mi$), and mass loss rate $\dot{M}$.  Its current
mass can be estimated from its integrated luminosity and/or kinematics, while
$\Mi$ and $\dot{M}$ require understanding of its orbital properties.

Here we adopt \textsc{emacss} in an iterative way to obtain $\Mi$ that correctly
reproduces $M$ given its orbital properties. However, we need two missing
ingredients, the cluster mass-loss history on eccentric orbits and its current
mass-to-light ratio ($\Upsilon$).

\comm{The version of \textsc{emacss} we use} is unable to evolve a cluster on an
eccentric orbit. To overcome this, we evolve the cluster on a circular orbit at
a Galactocentric radius $R_{\rm G,eqv}$, where the cluster has the same
total lifetime.  \comm{This is justified by the results of
\citet{2016MNRAS.455..596C} who  showed that the evolution of $M$ and $\rh$ are
similar for clusters on orbits with different eccentricities, but with the same
dissolution time.} This radius is given by \citep{Baumgardt03}

\begin{equation}
    \label{eq:eqvrad}
    R_{\rm G,eqv} = \rapo(1-\epsilon),
\end{equation}
where $\rapo$ is the apocentre radius and $\epsilon$ is the orbital
eccentricity.  These parameters are obtained by the orbit integration procedure
outlined in the section \ref{sec:orbit}.

With an approximation for the mass loss in eccentric orbits at hand, we now need
a prescription for $\Upsilon$. Due to preferential loss of low-mass stars and
stellar evolution, $\Upsilon$ evolves with time. Based on the work of
\citet{Anders09} we model $\Upsilon_V$, i.e. the mass-to-light ratio using the
luminosity in the $V$-band, as a linear function of \comm{the remaining mass in
the cluster}

\begin{equation}
    \label{eq:mtl}
    \Upsilon_{V} = 0.8 + 1.2\mud .
\end{equation}

Equation (\ref{eq:mtl}) gives $\Upsilon_V = 2$ if $\mud=1$, which is consistent
with a \citet{Kroupa01} IMF evolved to an age of 12\,\gyr\ and
$-2\lesssim\feh\lesssim-1$
\citep[e.g.][]{2003MNRAS.344.1000B,2010ApJ...712..833C}, while
$\Upsilon_V\simeq0.8$ if the remaining mass fraction approaches $\mud\simeq0$.
Note that in the $N$-body models of \citet{Anders09} $\Upsilon_V$ increases
again  near dissolution, because then the GC consists of predominantly dark
remnants, but this phase is short lasting hence we ignore it \citep[but we note
that NGC\,6535 may be in this phase,][]{2015ApJ...815...86H}. \comm{Since
\citet{Anders09} evolves an aged IMF, their results isolate dynamical effects
only.} \textsc{emacss}, \comm{however, accounts for stellar evolution during the
evolution of the cluster. In order to consistently use the \citet{Anders09}
relation, we assume the following relation $\mu = 2\,\mud$, which considers a
50\% mass-loss due to stellar evolution. The assumption is reasonable for
cluster older than $\sim$10 Gyr, which is the case for most GCs in the MW}

We proceed to use \textsc{emacss} to obtain the cluster initial mass. This is
done by minimizing the absolute value of the difference between \textsc{emacss}
predicted present-day mass and the cluster observed mass, where the observed
luminosity is converted to mass using Eq. \ref{eq:mtl}. The minimization
procedure uses the Nelder-Mead simplex method with a tolerance of 1 $\msun$. 

{\sc emacss} assumes a logarithmic Galactic potential, hence the mass evolution
is most accurate where the rotation curve of the Galaxy is flat. In the
innermost regions of the Galaxy ($<$ 3 kpc) {\sc emacss} will overestimate the
mass loss, providing only a lower limit on the cluster remaining mass fraction.
Clusters in the inner Galaxy will probe only a small fraction of the Galaxy
volume, as well as having tidal tails that are more easily disrupted due to
chaotic orbits and time-dependent potential from the bar \citep{APW16}, hence
our approximation holds for most GCs of interest.

\subsection{Orbit integration}
\label{sec:orbit}

The literature is rich regarding the measurement of GC proper motions (PMs), and
for about 2/3 of the known cluster there is a measurement available. For almost
all clusters a radial velocity measurement is available. However, both
these measurements are very heterogeneous.

We compiled a set of GCs from literature that have measured PMs.
The sample was further expanded by using UCAC-4 absolute PMs by
\citet{Dambis06}, however, the uncertainties in this catalogue can be
significant. The complete sample analysed here is shown in Table 1 where
appropriate references are given for each individual object.

We carry out orbit integration using the popular galactic dynamics code
\textsc{galpy} \citep{Bovy15}. The Galaxy potential used is the
\textsc{MWPotential2014} which is composed of three components: a power-law with
an exponential cut-off for the bulge, a \citet{MNdisk} disk, and a \citet*{NFW}
halo. We set the Sun's distance to the Galaxy centre to R$_{\odot}=8.3$ kpc, its
vertical offset to z$_{\odot}$ = 24.2 pc, and its circular velocity to $V_{\rm
c}$ = 233 km/s. The Solar reflex motion components adopted are the ones derived
by \citet{reflex}. We use the Dormand-Prince integration method \citep{DOPR}.
For the purpose of inferring orbital parameters, we integrate for 6 Gyr.

To propagate the uncertainties in position and velocity of each cluster we
perform 1500 orbit realisations for each cluster. In each realisation we sample
the values of PM from a normal distribution with a centre and dispersion given
by the literature value and uncertainty respectively. We also assume a 5\%
uncertainty on the heliocentric distance to the cluster, we find that this
assumption better encompasses the heterogeneous methodology for determining
cluster distances \comm{(the uncertainty in distance is propagated to the total
luminosity, as it depends on the distance modulus)}. We take the distance values
and integrated V-band magnitude from the  2010 version of the Harris catalog
\citep{Harris96, 2010arXiv1012.3224H}. At each realisation, the cluster initial
mass, current mass and remaining mass-fraction is computed according to the
recipe presented in this section. \comm{In addition to sampling the velocity and
position uncertainties, we also sample from the cluster age and associated
uncertainty. We take the ages from \citet{vdb13}, and for cluster where no age
is available we assume the mean age (11.9 Gyr) and dispersion (0.8 Gyr) of the
full sample. Although there are age determination for a larger sample spread
across the literature, we choose to use the largest one available in order to
avoid any methodology biases.}

With 1500 orbit samples we assume that the most representative value of $\rapo$,
$\rperi$, $\epsilon$, $\mu$, and $M_{{\rm i},5}$ is given by the median of the
distribution, while its uncertainty is given by the median absolute deviation
(MAD). Our choice of median and MAD is motivated by the fact that the median is
more sensitive to the asymmetries in the distribution while the MAD is less
sensitive to outliers. By comparison with the more traditional mean and standard
deviation we find that this approach gives more consistent results that are more
representative of all clusters in our sample, \comm{in the sense that objects
with very broad and asymmetric distributions have a median value that is more
representative of the full sample. While clusters with double peaked
distributions do not have a single peak favoured. This produces somewhat large
MAD values for some clusters, however these will certainly encompass the most
probable regions of parameter space.}

\begin{figure}
\centering
\includegraphics[width=0.45\textwidth]{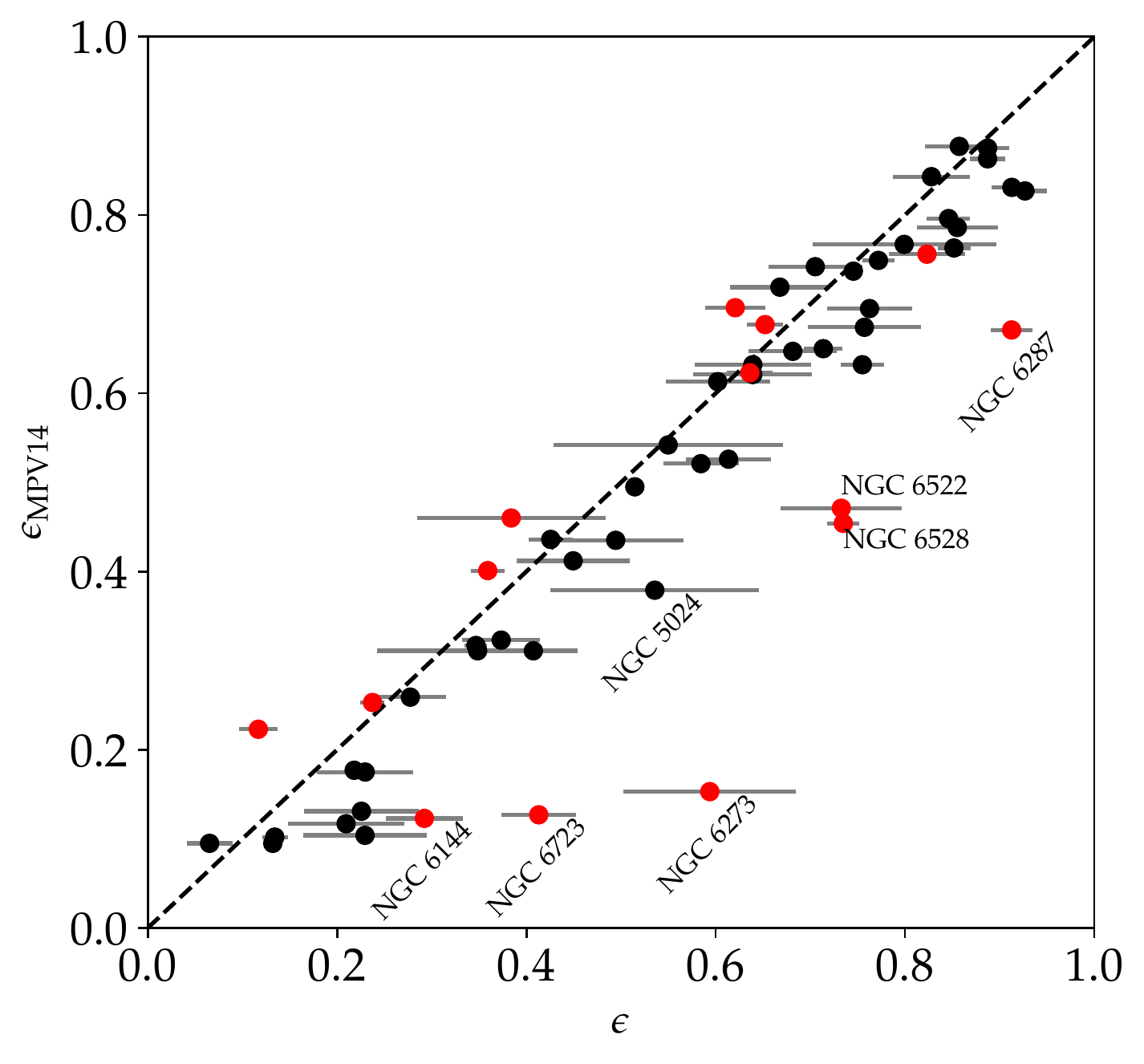}
    \caption{Comparison of the eccentricity obtained in this work ($x$-axis)
    with the ones obtained by \citep{mpv14} using an axisymmetric potential
    ($y$-axis).  Error-bars show the uncertainty in our measurements.
    \comm{Cluster in red have their current position within 3 kpc from the
    Galaxy centre, we indicate the names of those which deviate significantly
    from the identity.}}
\label{fig:ecc_comp}
\end{figure}

To check the consistency of our orbit integration, we compare our eccentricity
values with those in the axisymmetric case in \citet{mpv14}. The comparison is
restricted to clusters in common with their sample that use the same PM values,
since we adopt more recent PM measurements for some of the clusters.  In
Fig.~\ref{fig:ecc_comp} we show the comparison with \citet{mpv14} ($y$-axis) and
our eccentricity determination ($x$-axis). We flag outliers by selecting objects
that lie more than 2 times the typical eccentricity uncertainty. The bulk of the
cluster lie along the identity line within the uncertainties. The outliers are
clusters for which the orbit is strongly influenced by the bulge potential. Our
central potential is slightly different than the one used by \citet{mpv14} which
could be the source of the discrepancy.

\begin{figure*}
\centering
\includegraphics[width=0.85\textwidth]{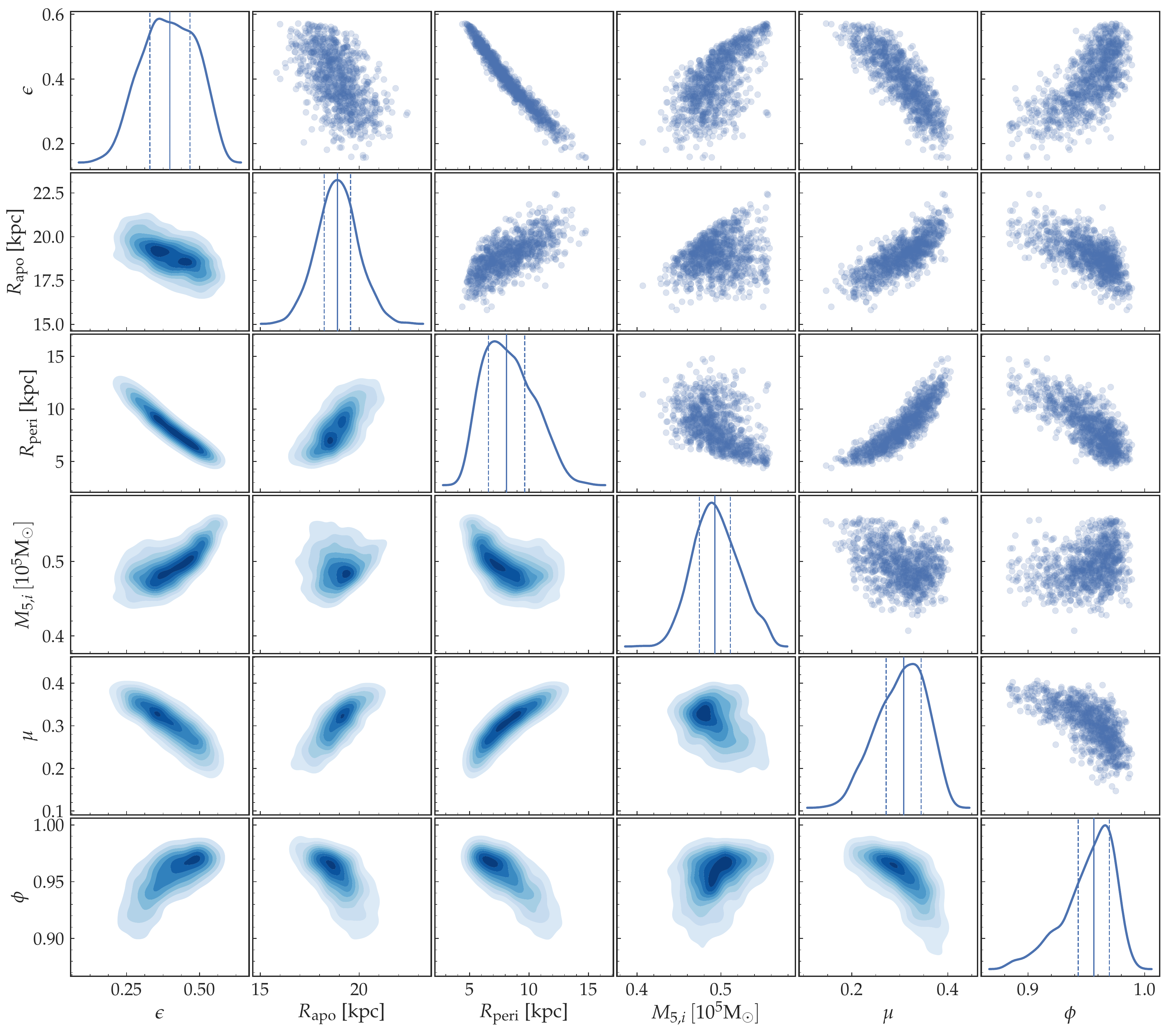}
    \caption{\comm{Summary plot} showing the relation among all parameters explored in
    the orbit integration of Palomar 5 as an example. Panels below the diagonal
    show the 2D kernel density estimator for the samples, panels above the
    diagonal show the individual samples, and the diagonal shows the 1D kernel
    estimator with median (solid line) and MAD (dashed line) indicated.} 
\label{fig:p5samples}
\end{figure*}

In Fig.~\ref{fig:p5samples} we show one example of the orbital parameter
exploration for Palomar 5. The diagonal plots show the 1D kernel density
estimator (KDE) for each of the parameters sampled. The panels below the
diagonal show the 2D KDE for each pair of parameters and the panels above show
each individual sample. We also include the orbital phase $\phi$ which is
computed using $\phi = (\rg - \rperi)/(\rapo- \rperi)$.

In order to get a complete picture of the Galactic GC system, we assume that GCs
without PM measurements have an  isotropic velocity distribution (hereby
referred as the noPM sample). \citet{vandenBosch99} show that in a logarithmic
halo, a tracer population with density $\rho\propto r^{-\alpha}$ has a velocity
dispersion $\sigma_{\rm GC}= V_{\rm c}/\sqrt{\alpha}$, and for the GCs
$\alpha\simeq3.5$ \citep{Harris76}. These clusters form about 1/3 of the sample
and their orbits were sampled in the same fashion as before. However, we sample
the velocities isotropically from a normal distribution with dispersion
$\sigma_{\rm GC}$. Cluster that fall in this category are indicated in Table 1.
\comm{In addition to the clusters without PM, we moved two objects to the noPM
sample, these are Terzan 8 and IC 4499. Their literature PM yielded unbound
orbits, however, given that these come from the less accurate \citet{Dambis06}
sample, we assume that this is due to measurement errors.  Hence, had their PM
removed from Table 1, and their orbits where sampled from the isotropic velocity
distribution. We note that the clusters Pal 3, Rup 106, NGC 5634, IC 1257, and
NGC 6426 have extremely high $\rapo$, which is likely due to inaccuracies in
their PM measurement. However, since we were able to recover bound orbits, we
choose to keep these objects in our sample.  We also include a derivation of the
Jacobi radius ($\rj$) for each cluster at their current position. We use the
\citet{King62} formula and use our derived present-day mass to propage the
uncertainty to $\rj$.}

\begin{figure}
\centering
\includegraphics[width=0.45\textwidth]{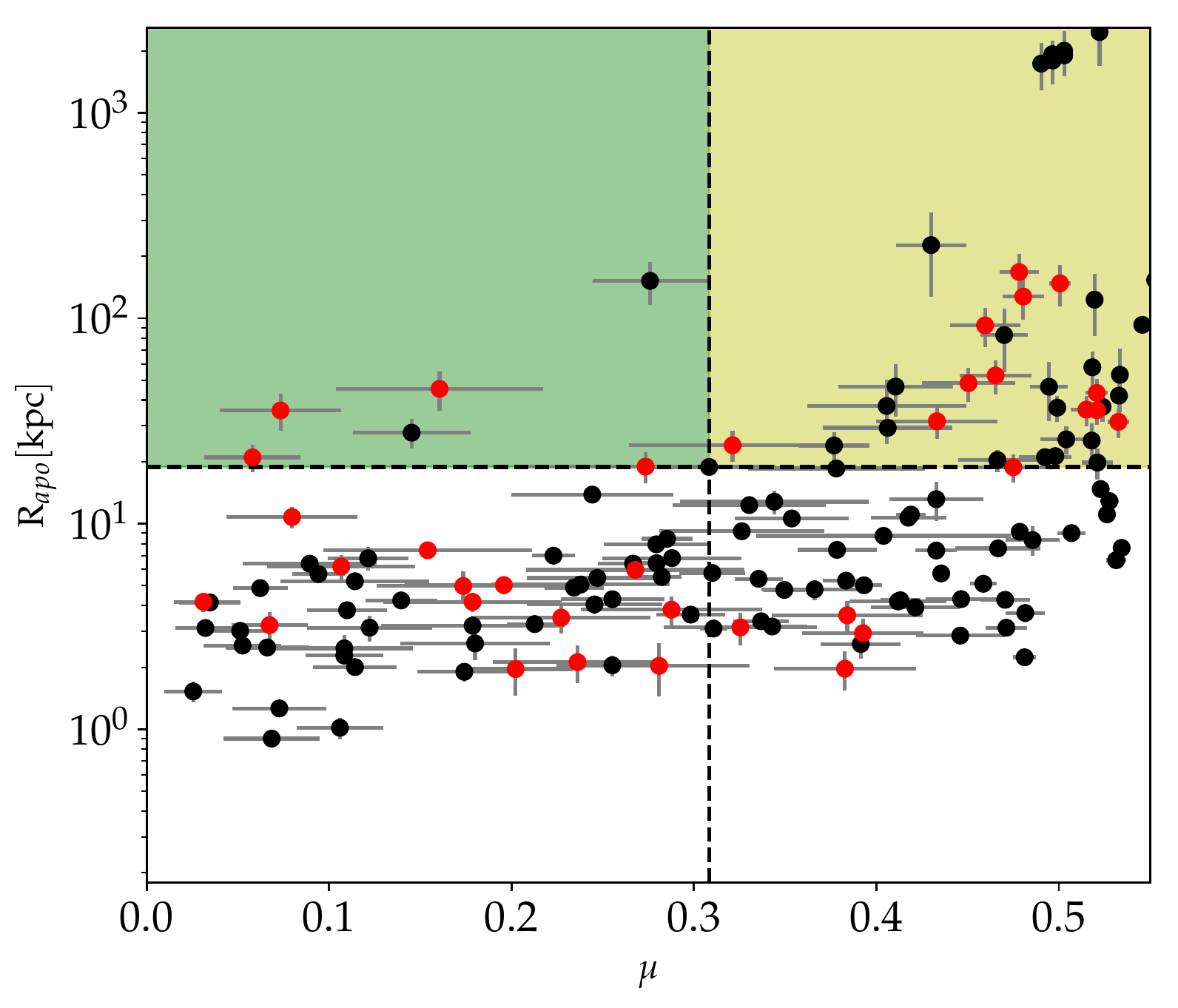}
    \caption{Distribution of $\mu$ and $\rapo$. Errorbars were computed using
    the MAD estimator. Red circles are cluster  from the noPM sample. The
    green shaded region show cluster that have both orbital parameters more
    extreme than Palomar 5, while the yellow shaded region show clusters that
    have a $\rapo$ larger than Palomar 5. The clusters in the green region are
    AM 4, Palomar 1, Palomar 7 (IC 1276), and Whiting 1. Note that the maximum
    $\mu\simeq0.55$ is because all GCs have lost about 45\% of their initial
    mass as the result of stellar evolution.}
\label{fig:mu}
\end{figure}

In Fig.~\ref{fig:mu} we show the resulting distributions of $\rapo$ and $\mu$.
We note that GCs closer to dissolution (low $\mu$) tend to have smaller $\rapo$,
{\i.e. clusters closer to the Galactic centre where the tidal field of the bulge
is strong}, while clusters with higher $\mu$ span a wider range of $\rapo$.
{Cluster also have a maximum $\mu \sim 0.55$ which is due to the fast mass loss
from stellar evolution that remove about half of $\Mi$ at an age of \comm{$\sim$
12 Gyr.}}  We mark regions of  parameter space that are more extreme than what
is observed for Palomar 5 (i.e. closer to disruption and/or with a larger
$\rapo$). \comm{This selection is motivated by the usefulness of streams as
tracers of the potential beyond 20 kpc.} Based solely on these parameters and
$\mu$, we expect clusters in the \emph{green} region of the parameters space to
be good candidates for tidal tail search. These cluster are AM4, Palomar 1,
Palomar 7 (IC 1276), and Whiting 1.  Of these clusters, only Palomar 1 has
reported tidal tails \citep{pal1}.  These clusters are also at large
heliocentric distances which makes obtaining PMs more difficult and only Palomar
7 has measured PMs. The noPM sample cluster mentioned above, despite having low
uncertainties in $\mu$ and $\rapo$, may suffer from additional biases and their
location in this plane is not as reliable as for the PM sample, hence our
estimates may not hold once PM becomes available for these objects.

We note that our cluster mass estimates are based on the assumption of a static
MW potential, in the sense that all cluster orbit the same potential -- and on a
given orbit -- during their whole lifespan. In comparison to the real orbital
history of a cluster, our estimates may yield discrepant results. If we would
account for the secular growth of the Milky Way, we would find $\mu$ values that
are slightly higher because the average tidal field experienced in the past is
weaker \citep{Renaud15}. Accreted clusters from minor or major merger events
could have been in radically different environments for a significant fraction
of time \citep*{Renaud17}. This would ultimately leave an imprint on the cluster
evolution such as a present day mass function (PDFM) that is not consistent with
is current orbit.

With the  properties of the orbits and $\mu$ for all GCs, we now discuss the
visibility of streams in the next section.

\section{The visibility of a stream}
\label{sec:visibility}

We have established that the cluster dissolution stage is a major factor in the
typical mass of the stars in the stream. Nonetheless, it is known that streams
are in their most densely packed stage near $\rapo$. So we expect that the most
easily detectable streams are those from clusters with low $\mu$ (i.e. near
dissolution) and high $\phi$ (i.e. near $\rapo$).

\begin{figure}
\centering
\includegraphics[width=0.45\textwidth]{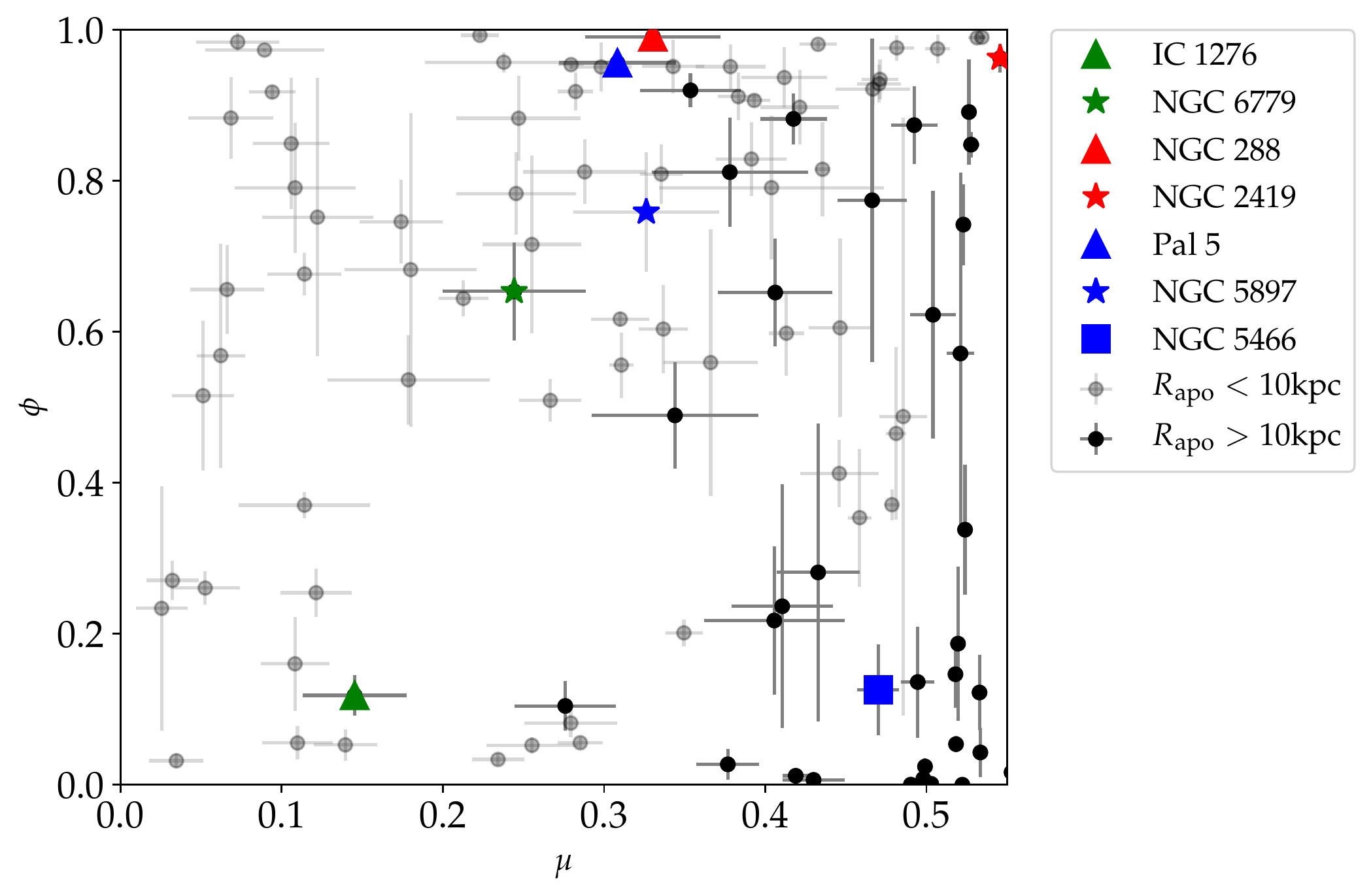}
\caption{Remaining mass $\mu$ against orbital phase $\phi$ (1 at apocentre),
    excluding the noPM sample. In grey we show cluster with $\rapo<10$ kpc, and
    in black cluster with $\rapo > 10$ kpc. Error-bars are derived from the MAD
    estimator. Cluster with detected tidal tails are marked in blue. From the
cluster with $\rapo > 10$ kpc we highlight the two closest to disruption (green
symbols) and the two closest to apocentre (red symbols).}
\label{fig:best}
\end{figure}

In Fig.~\ref{fig:best} we show the distribution of $\mu$ and $\phi$ for all
clusters, except those in the noPM sample, since these have unreliable values of
$\phi$. We split the sample into clusters that have $\rapo>10$ and $<10$ kpc. We
highlight Palomar 5 \citep{Odenkirchen01}, NGC 5897 (Adrian Price-Whelan private
communication), and NGC 5466 \citep{Belokurov06, Grillmair06b} since these
clusters have tidal tails detected in SDSS or Pan-STARRS. \comm{Recently, tidal
tails have been found around the clusters NGC 7492, Eridanus, and Pal
15 \citep{2017ApJ...841L..23N, 2017ApJ...840L..25M}, however, these can not be
used in our comparison as no PM measurements are available.} For those with
$\rapo> 10$ kpc we show the two lowest $\mu$ values (green) and the two highest
$\phi$ values (red). These objects are high-probability clusters for detection
of streams considering SDSS-like photometric performance. 

The cluster NGC 5466 seems to be an exception, in the sense that it should not
have easily detectable tails based on its relatively high $\mu$ and being near
$\rperi$, but \citet{Belokurov06} detected a 4$^{\circ}$ tail around this
cluster using SDSS data, followed by \citet{Grillmair06b} who detected it up to
45$\deg$. We point out that NGC 5466 is a cluster that has an PDMF that is
depleted in low-mass stars \citep{Sollima17} and \comm{it has been suggested by
\citet{Webb17} that this cluster is not mass segregated. According to our
prediction, NGC 5466 should have a pristine PDMF since $\mu\simeq0.5$.
Interestingly, the cluster is also on a retrograde, high eccentricity orbit
\citep{Forbes10}}, which is a strong indication for it being an accreted cluster.
Here we strengthen the suspicion about its accreted origin, since it has an PDMF
inconsistent with its orbital history which could be the effect of a stronger
tidal field in the past, perhaps from its original host galaxy.

As shown in Fig.~\ref{fig:best}, the two clusters that are closest to the
optimal detectability conditions (low $\mu$ and high $\phi$) and have
$\rapo>10\,\kpc$ are NGC 288 and NGC 6341 (M92). The former has weak signs of
tidal tails \citep{Grillmair04} in shallow IR data, while the latter has signs
of extra-tidal structure \citep{Lee03}.  We also highlight two clusters that are
close to disruption and have $\rapo > 10$ kpc, which are IC 1276, also known as
Palomar 7, and NGC 6779 (M56).  However, both clusters are quite close to the
bulge in projection ($|b| < 6^{\circ}$), which will likely introduce
differential reddening effects and a high surface density of field stars which
can make stream detection challenging.

The criteria for being an optimal target for finding tidal tails appears to
yield consistent predictions, at least based on the present literature. We
stress, however, that each cluster is a particular case in the sense that its
distance, field contamination, and orbit projection should be considered since
these factors may produce non-trivial observing conditions. For instance, the
cluster M92 is an optimal candidate, however its proximity to the Sun will lower
the projected surface density. With such complications in mind, our predictions
should be taken as general constraints that can be used as guidelines for future
observations and follow-ups.

Finally, this framework can be used to predict not only the spectral properties
of stream stars (colours and magnitudes), but also their distribution in PM
space of stream stars. With the velocity information, disentangling field stars
from stream stars will be much more efficient than current photometry-based
methods. This may allow the study of streams in the inner Galaxy, provided
extinction is not an issue.

\section{The case of Palomar 5}

Using the information about the remaining mass in a given cluster and its
orbital history we are able to track the masses of the particles that leave the
cluster. However, we still need a realistic way of simulating the ejection
mechanism and produce a tidal tail.

In a cold stream self-gravity is not important.  This  greatly simplifies the
stream simulation in the sense that one only needs to integrate individual
orbits of particles being \emph{sprayed} off a progenitor in a static potential
to produce a realistic stream. This class of codes became known as 
spray-particle methods and has been explored in many successful applications
\citep{Kuepper12, Fardal15, Sesar15, Erkal16, Amorisco15, Bonaca14}. Despite the
simplicity, this method still requires prescriptions for how particles are
ejected from the progenitor. We refer to \citet{Fardal15} for a comprehensive
comparison of methods.

\begin{figure*}
\centering
\includegraphics[width=0.95\textwidth]{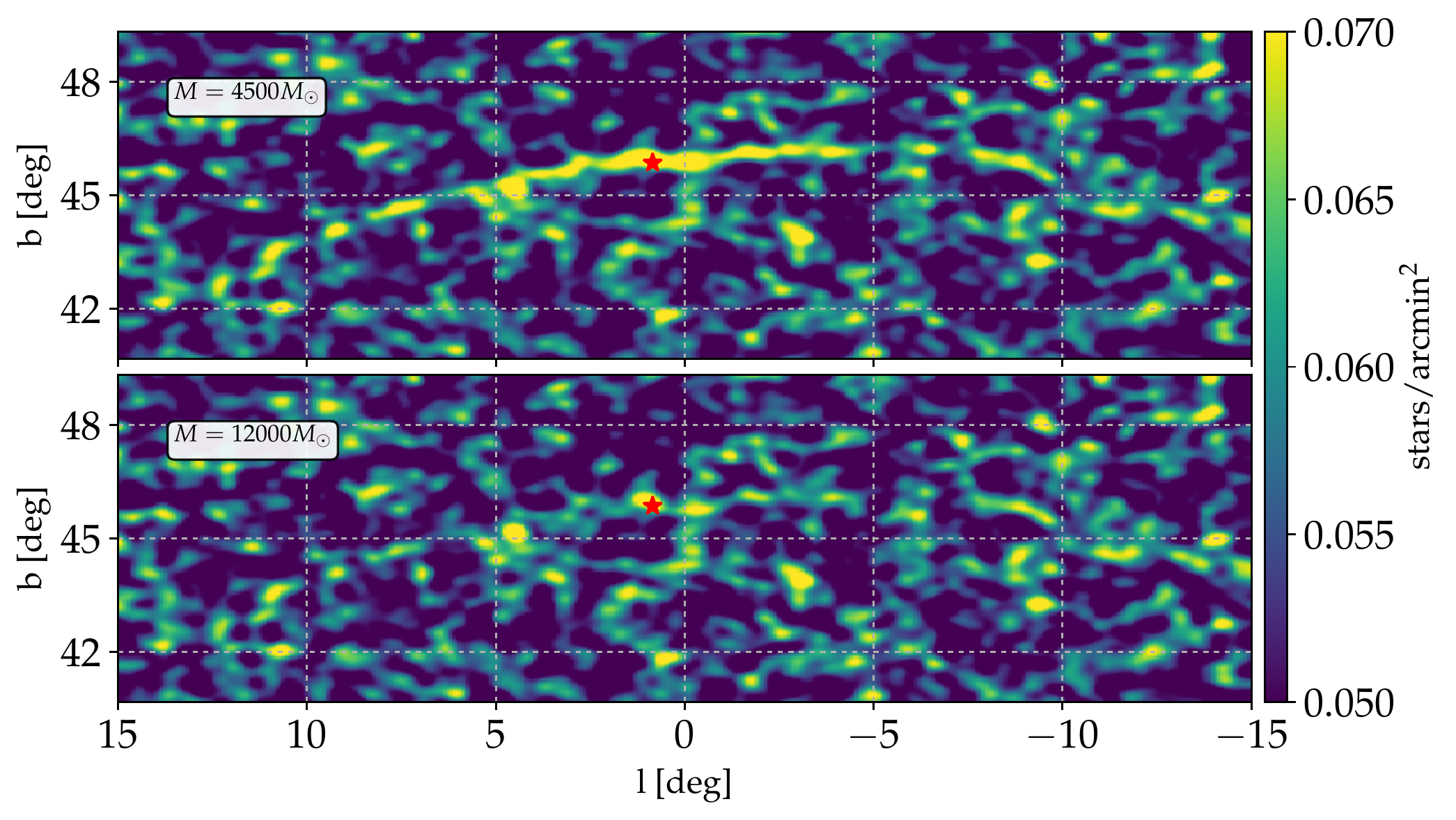}
    \caption{Two simulations of Palomar 5 tidal tails. The top panel show a
    Palomar 5 that is close to its estimated present-day mass
    \citep{Odenkirchen01} while the bottom panel it is presently $\sim 3$ times
    more massive. The red star symbol marks the position of the cluster.} 
\label{fig:p5onsky}
\end{figure*}

In this work, we adopt the spray-particle implementation by \citet{Kuepper12}.
The escape velocity of stars was modelled based on an extensive study of
$N$-body simulations in tidal fields presented by \citet*{Claydon16}.  These
authors studied so-called `potential escapers' in their simulations, which are
stars that are energetically unbound, but still associated with the cluster.
From this, and theoretical arguments, they find that the velocity dispersion of
these stars near the Jacobi radius ($\rj$) can be approximated by 

\begin{align}
\sigma_{\rm J} \simeq 0.9\,\kms\left(\frac{M}{10^5\,\msun}\right)^{1/4}\left(\frac{\rg}{5\,\kpc}\right)^{-1/3}.
\end{align}

The approximation above is valid for circular orbits. In order to use it for
eccentric orbits we find the circular orbit that has the same life-time given by
equation~(\ref{eq:eqvrad}).

The release mechanism adopted here assumes a Maxwellian velocity distribution
for the escapers with the characteristic velocity scale  equal to $\sigma_{\rm
J}$. Particles are ejected from the Lagrange points plus a random offset given
by a $0.3\rj$ dispersion normal distribution. The extra offset in position is
required to match $N$-body simulations \citep{Kuepper12, Lane12, Bonaca14,
Fardal15, Pearson15}.

Each particle from the simulated tidal stream represents an ensemble of stars
with a known MF (i.e. given by the moment the particle escaped the cluster and
the algorithm of section~\ref{sec:evolve_mf}). Combining this information with a
mass-luminosity relation we can produce an expected luminosity function (LF) for
each particle in the simulation.  For Palomar 5 we choose an 11 Gyr isochrone
\citep{Bressan12} \footnote{\url{http://stev.oapd.inaf.it/cgi-bin/cmd_2.8}} with
a metallicity of $\feh= -1.4$ \citep{Koch04}.  To simulate SDSS-like photometry
we use a faint magnitude limit in the $g$-band of 22.5. Each tail particle has
its associated EMF integrated from the turn-off mass to the mass that
corresponds to the chosen magnitude limit, taking into account the heliocentric
distance to each individual particle. The result is a list of particles with
on-sky positions and a weight that is proportional to the number of observable
stars it represents.  We produce density maps by counting stars in bins on the
sky and dividing by the solid angle of each bin.

To illustrate the effects of the MF evolution on the detectability of Palomar
5's stream we perform two simulations. Both in the same orbit, but with one that
is consistent with the lower limit of its present-day mass of 4500 $\msun$
\citep{Odenkirchen02} and another which has a current mass that is roughly 3
times higher (12000 $\msun$). To artificially increase its current mass, we
increase its observed luminosity and run through our $\mu$ estimate method
described in Section \ref{sec:orbit}.

Tidal tails are sparse structures and to estimate the detectability of a given
stream one must consider the effect of overlapping field stars. To estimate the
contribution of field stars we analyse a 4 deg$^2$ patch of the sky observed by
SDSS. We filter stars in colour and magnitude according  to the best-fit
isochrone in literature \citep{Koch04}. We select stars with colours $(g-r)$
$\pm 1.5 \sigma_{g-r} (g)$ away from the isochrone, where $\sigma_{g-r}(g)$ is the
typical colour uncertainty at a given magnitude $g$. We apply the same $g$-band
cut-off at $g=22.5$ as before. In the patch of the sky analysed this yields a mean
density of 0.058 stars/arcmin$^2$ which was used to generate a homogeneous
distribution in the region around the Palomar 5. This approach does not account
for variations in the number of field stars with e.g. galactic latitude, however
it is representative of the field stars density in the vicinity of the cluster.
It is also a first order approximation of the traditional matched-filter
approach used to detect streams \citep{Odenkirchen01,Balbinot11}, in the sense
that it selects only stars with colours consistent with the cluster population.

\begin{figure*}
\centering
\includegraphics[width=0.95\textwidth]{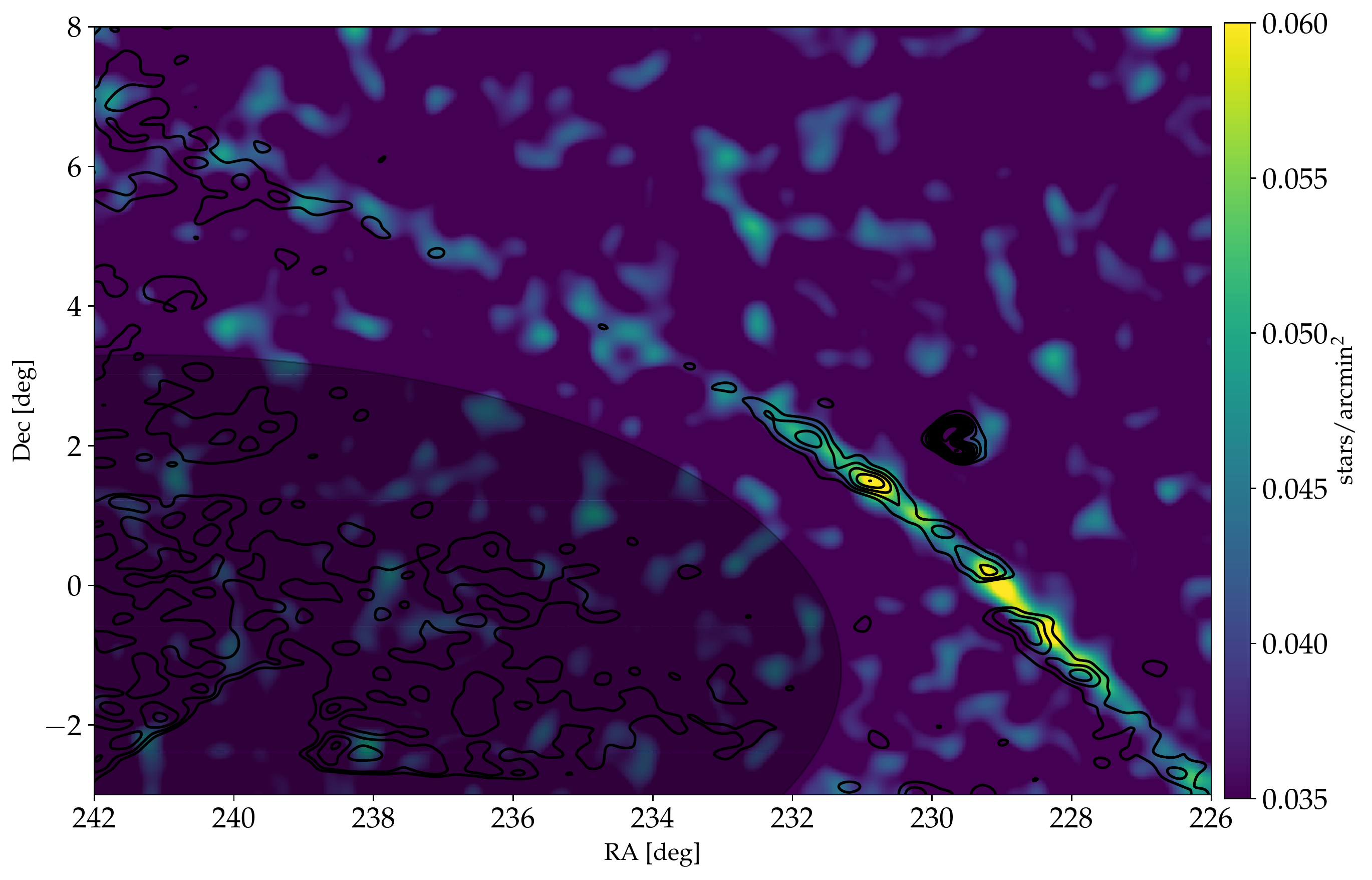}
    \caption{The same simulation of Palomar 5 as shown in the top panel of
    Fig.~\ref{fig:p5onsky} where the background has been lowered to a level
    consistent with the decontamination efficiency of the matched-filter
    technique. Contours are the 1,2,3,4, and 5 $\sigma$ confidence levels of the
    SDSS detection, based on the work of \citet{Balbinot11, Kuepper15}. The
    shaded region delimits a portion of the sky where contamination by the bulge
    becomes important and the SDSS contours are unreliable.  The contour blob at
    $(\alpha, \delta) \simeq (230, 2)$ is the foreground cluster M5.} 
\label{fig:p5SDSS}
\end{figure*}

In Fig.~\ref{fig:p5onsky} we show the predicted density of stars for the two
simulated Palomar 5 tails. The present-day mass is indicated in each panel.
Field stars were added as described above and both maps are smoothed using a
Gaussian kernel with 0.075 deg dispersion. We notice that the model with the
mass in agreement with observations shows more prominent tails that are clearly
visible in the range $-5^{\circ} < l < 5^{\circ}$. Also the trailing tail (positive $l$) appears
slightly more extended, which is also confirmed by Pan-STARRS observations
\citep{Bernard16}.  The average density of the stream plus background is of
0.069 stars/arcmin$^2$, which is in agreement with SDSS data
\citep{Odenkirchen01}. We note, however, that some of the fainter stream
features are somewhat dependent on the random background realisation. The more
massive simulation has a mean density $\sim$ 20\% lower than the normal mass
one, even though its mass loss rate is $\sim$ 30\% higher. The density decrease
makes most of the stream fall below the background level. The different density
can be explained by the median mass of stream stars: the stars in the realistic
Pal 5 stream simulation have a median mass of 0.3 $\msun$, while only 0.22
$\msun$ in the massive one.

Finally, we compare our stream prediction to SDSS data. We use the
matched-filter density map from \citet{Balbinot11,Kuepper15} which follows
closely the methodology outlined by \citet{Odenkirchen01}, the discovery paper.
For a fair comparison between SDSS matched-filtered data and our simulated
stream plus background we must take into account that the matched-filter has a
higher efficiency in classifying stream stars. We assume that field stars that
fall within the isochrone defined colour-magnitude mask -- discussed above --
are discarded with a 50\% efficiency. Effectively, this is the equivalent of
lowering the background level by 50\%.

In Fig.~\ref{fig:p5SDSS} we show our matched-filter equivalent density map for
the 4500 $\msun$ simulation. The overlaid contours are from \citet{Balbinot11}
and \citet{Kuepper15}, where we show the detection confidence levels from 1 to 5
$\sigma$. The shaded region marks regions where the SDSS footprint becomes
patchy and/or too close to the bulge causing spurious detections. Notice that
the modelled stream is in very good agreement with the observational data.
Most notably it reproduces the length on the stream main body and some of the
over/under dense regions. We note that the length is reproduced even if we
assume a matched-filter decontamination efficiency up to 80\%. Despite the very
good agreement, we stress that some of the features observed are dependent on
the background realization, indicating that these are close to the noise level.
As pointed out by \citet{Thomas16} some of the features observed in SDSS do not
stand the trial of deeper photometric data.

Even though the comparison with SDSS show great potential, we leave a more
detailed analysis of Palomar 5 for future works where we explore in more detail
its stream properties. For instance, we do not explore different methods of
stream generation, $N$-body simulations or a more detailed projection onto
observational space (e.g.completeness, source miss-classification, etc). The goal
of this work is to illustrate the nuances generated by collisional dynamics and
their overall impact on stream properties.

\section{Conclusion}

We present a method for taking into consideration the internal evolution of a
cluster due to collisional dynamics on the formation of cold streams. This
method uses the code \textsc{emacss} \citep{Alexander14} to simulate the cluster
main properties such as total bound mass, number of stars, and mass-loss rate.
We couple this information to a algorithm that evolves the MF based on
simplified version of the method presented in \citet{Lamers13}.

The median mass of the escaping stars is always below the median mass of the
bound stars, and the mass of escapers gradually increases as the evolution
proceeds. Our predictions \comm{for the evolution of the mass function of stars
and stellar remnants} are in good agreement with \comm{results of an} $N$-body
model.

Using position and velocity information for MW GCs we derive orbital parameters
and their uncertainties. These were coupled to a simple mass-loss prescription
to infer the cluster remaining and initial masses, effectively allowing us to
assess the current stage of dissolution of each cluster. We validate our orbit
integration by comparing it to the sample of \citet{mpv14}, which we find to be
in good agreement.

As a test case, we show a detailed simulation of the cluster Palomar 5. We use
the same spray particle code used in \citet{Kuepper15} with the additional
prescription presented in \citet{Claydon16} for the velocity dispersion of
escaper stars. We generate two streams, a `normal' one which is consistent with
the cluster present-day mass, and an `overweight' one which yields a cluster
that is 3 times more massive today. We find that the `normal' mass simulation
successfully predicts the surface density of stream stars as observed by SDSS,
as well as some of the features in the stream (asymmetry and some
over-densities). On the other hand, the `overweight' simulation yields a stream
that is below the SDSS detection limit throughout most of its extension.
\comm{The simulations presented here by no means intend to be an exhaustive
exploration of Palomar 5 stream structure. We suggest the works by
\citet{Pearson15}, \citet{Kuepper15}, and \citet{APW16} for a detailed treatment
of stream formation and structure in a variety of scenarios.}

We use Palomar 5, and other clusters with detected streams in SDSS or
Pan-STARRS, as a guideline to probe properties that enhance the probability of
stream detection. As demonstrated, most cluster with stream are close to
dissolution \comm{and/or have atypical PDMFs \citep{Webb17}}. We also find that these clusters
are more likely to be close to apocentre as well, a phase where stream stars are
in their most densely packed configuration. From the sample with PM we single
out the clusters NGC 288 and NGC 6341 (M92) as the next best candidates after
Palomar 5, to have detectable tidal tails. Indeed there is evidence in
literature for this \citep{Lee03, Grillmair04}. From the noPM sample, we
highlight the clusters AM4, Palomar 1, Palomar 7, and Whiting 1. These clusters
have $\rapo$ larger than 18 kpc and are closer to dissolution than Palomar 5.
Among these clusters, only Palomar 1 has evidence for tidal tails \citep{pal1},
although wide-field data is lacking for the other three. 

\citet{1984ApJ...277..470P} suggest that GCs form in small dark
matter halos, and if these are still surrounding the cluster, they prevent stars
from escaping.  \citet{1996ApJ...461L..13M} argues that the presence of tidal
tails near some GCs and dwarf galaxies implies that these objects can not have a
massive  dark matter halo. This argument could be turned around to use the
absence of tails near massive GCs \citep[e.g.][]{Kuzma16} as evidence for the
presence of a dark matter halo. Our results provide a more natural explanation
for the absence of tidal streams near massive GCs, despite their higher mass
loss rate (at a given orbit).

The overall distribution of cluster orbital properties shows how rare objects
like Palomar 5 are, and certainly provide a dire perspective for the use of cold
streams to probe the halo beyond 18 kpc. However, progenitor-less streams
outnumber those associated with GCs \citep[for recent
discoveries see][]{Bernard16}, offering hope for probing the halo at
larger distances. Although, the number of cluster close to disruption in the
outer halo is expected to be low. \citet{Contenta17} recently showed that
the number of cluster close to disruption at Galactocentric radius between
20 and 150 kpc is $3.3^{+7.7}_{-1.6}$. However, there may be a larger
population of progenitor-less streams, if the orbital phase mixing time is
long. Also, some of the  low-mass cluster in the outer halo show signs of
streams, such as Willman 1 \citep{Willman06}, Balbinot 1 \citep{Bernard16}, 
Segue 1 \citep{segue1tail}, \comm{and Hercules \citep{Kuepper17}}.

Upcoming surveys such as the Large Synoptic Survey Telescope
\citep[LSST][]{LSST} will greatly increase sky coverage and photometric depth,
potentially revealing many new streams. However, greater photometric depth comes
at the cost of an increase density of field stars {and background galaxies}.
Next generation detection methods must account for this and devise ways of
better modelling the stream as well as background. The addition of velocity
information will certainly help disentangling stream stars from field stars, and
the next data release of Gaia will provide a unique test case for new methods.

\section*{Acknowledgments}

We thank Holger Baumgardt for kindly providing NBODY model used in this work and
Andreas K{\"u}pper for providing the streakline code. MG thanks the Royal
Society for financial support (University Research Fellowship) and both authors
acknowledge financial support from the European Research Council
(ERC-StG-335936, CLUSTERS). Both authors thank Justin Read and Jeremy Webb for
interesting discussions and suggestions.

SDSS-III is managed by the Astrophysical Research Consortium for the
Participating Institutions of the SDSS-III Collaboration including the
University of Arizona, the Brazilian Participation Group, Brookhaven National
Laboratory, Carnegie Mellon University, University of Florida, the French
Participation Group, the German Participation Group, Harvard University, the
Instituto de Astrofisica de Canarias, the Michigan State/Notre Dame/JINA
Participation Group, Johns Hopkins University, Lawrence Berkeley National
Laboratory, Max Planck Institute for Astrophysics, Max Planck Institute for
Extraterrestrial Physics, New Mexico State University, New York University, Ohio
State University, Pennsylvania State University, University of Portsmouth,
Princeton University, the Spanish Participation Group, University of Tokyo,
University of Utah, Vanderbilt University, University of Virginia, University of
Washington, and Yale University.

\newpage
\begin{landscape}
\begin{table}
\footnotesize
\begin{tabular}{lccccccccccc}
\hline
Name & $\mu_{\alpha}$ & $\mu_{\delta}$ & V$_{\rm los}$ & R$_{apo}$ & R$_{peri}$ & ecc & M$_{i}$ & $\mu$ & $\phi$ & $r_J$ & Refs \\\\
 & mas/yr & mas/yr & km$\,$ s$^{-1}$ & kpc & kpc & & $10^5$ M$_{\odot}$ & & & pc & \\
\hline
\hline
NGC 104 & 5.64$\pm{0.2}$ & -2.02$\pm{0.2}$ & -18.0$\pm{0.1}$ & 7.63$\pm{0.02}$ & 5.85$\pm{0.1}$ & 0.13$\pm{0.01}$ & 18.89$\pm{1.24}$ & 0.53$\pm{0.01}$ & 0.99$\pm{0.01}$ & 137.45$\pm{3.39}$&3 \\ 
NGC 288 & 4.67$\pm{0.42}$ & -5.62$\pm{0.23}$ & -45.4$\pm{0.2}$ & 12.32$\pm{0.21}$ & 2.72$\pm{0.59}$ & 0.64$\pm{0.06}$ & 2.05$\pm{0.08}$ & 0.33$\pm{0.04}$ & 0.99$\pm{0.01}$ & 76.43$\pm{4.02}$&1,3 \\ 
NGC 362 & 5.07$\pm{0.71}$ & -2.55$\pm{0.72}$ & 223.5$\pm{0.5}$ & 10.69$\pm{0.47}$ & 0.84$\pm{0.2}$ & 0.86$\pm{0.04}$ & 8.51$\pm{0.52}$ & 0.42$\pm{0.02}$ & 0.88$\pm{0.03}$ & 112.06$\pm{4.57}$&3 \\ 
Whiting 1* & -- & -- & -- & 46.22$\pm{10.17}$ & 20.64$\pm{6.36}$ & 0.41$\pm{0.11}$ & 0.06$\pm{0.01}$ & 0.17$\pm{0.05}$ & 0.5$\pm{0.31}$ & 37.56$\pm{2.28}$&-- \\ 
NGC 1261 & 1.33$\pm{0.89}$ & -3.06$\pm{1.06}$ & 68.2$\pm{4.6}$ & 25.54$\pm{3.74}$ & 5.72$\pm{2.17}$ & 0.65$\pm{0.06}$ & 4.41$\pm{0.27}$ & 0.5$\pm{0.01}$ & 0.63$\pm{0.16}$ & 146.38$\pm{6.31}$&8 \\ 
Pal 1* & -- & -- & -- & 21.72$\pm{3.85}$ & 9.7$\pm{3.4}$ & 0.41$\pm{0.12}$ & 0.13$\pm{0.04}$ & 0.06$\pm{0.03}$ & 0.6$\pm{0.29}$ & 22.59$\pm{2.13}$&-- \\ 
AM 1* & -- & -- & -- & 155.74$\pm{27.37}$ & 55.06$\pm{27.66}$ & 0.52$\pm{0.17}$ & 0.27$\pm{0.02}$ & 0.48$\pm{0.01}$ & 0.65$\pm{0.27}$ & 203.08$\pm{9.32}$&-- \\ 
Eridanus* & -- & -- & -- & 119.12$\pm{22.17}$ & 43.68$\pm{21.38}$ & 0.5$\pm{0.17}$ & 0.38$\pm{0.03}$ & 0.48$\pm{0.01}$ & 0.64$\pm{0.28}$ & 191.61$\pm{8.84}$&-- \\ 
Pal 2* & -- & -- & -- & 41.88$\pm{5.86}$ & 13.7$\pm{7.14}$ & 0.52$\pm{0.18}$ & 5.06$\pm{0.34}$ & 0.52$\pm{0.01}$ & 0.75$\pm{0.22}$ & 242.15$\pm{9.84}$&-- \\ 
NGC 1851 & 1.28$\pm{0.68}$ & 2.39$\pm{0.65}$ & 320.5$\pm{0.6}$ & 37.02$\pm{5.79}$ & 6.59$\pm{1.17}$ & 0.71$\pm{0.02}$ & 6.99$\pm{0.44}$ & 0.52$\pm{0.01}$ & 0.34$\pm{0.09}$ & 166.46$\pm{6.04}$&1,3 \\ 
NGC 1904 & 2.12$\pm{0.64}$ & -0.02$\pm{0.64}$ & 205.8$\pm{0.4}$ & 21.07$\pm{0.79}$ & 5.44$\pm{1.7}$ & 0.59$\pm{0.09}$ & 4.69$\pm{0.3}$ & 0.49$\pm{0.01}$ & 0.87$\pm{0.05}$ & 153.79$\pm{6.0}$&2,3 \\ 
NGC 2298 & 4.05$\pm{1.0}$ & -1.72$\pm{0.98}$ & 148.9$\pm{1.2}$ & 18.59$\pm{1.12}$ & 5.29$\pm{1.84}$ & 0.56$\pm{0.11}$ & 1.29$\pm{0.1}$ & 0.38$\pm{0.05}$ & 0.81$\pm{0.07}$ & 81.27$\pm{4.08}$&2,3 \\ 
NGC 2419 & -0.17$\pm{0.26}$ & -0.49$\pm{0.17}$ & -20.2$\pm{0.5}$ & 92.19$\pm{2.93}$ & 30.77$\pm{13.47}$ & 0.5$\pm{0.16}$ & 18.88$\pm{1.21}$ & 0.55$\pm{0.01}$ & 0.97$\pm{0.01}$ & 719.16$\pm{30.56}$&8 \\ 
Pyxis* & -- & -- & -- & 50.72$\pm{8.33}$ & 18.25$\pm{8.51}$ & 0.51$\pm{0.15}$ & 0.68$\pm{0.05}$ & 0.46$\pm{0.02}$ & 0.69$\pm{0.25}$ & 132.14$\pm{6.52}$&-- \\ 
NGC 2808 & 0.58$\pm{0.45}$ & 2.06$\pm{0.46}$ & 101.6$\pm{0.7}$ & 12.89$\pm{0.37}$ & 2.42$\pm{0.38}$ & 0.68$\pm{0.05}$ & 18.52$\pm{1.17}$ & 0.53$\pm{0.01}$ & 0.85$\pm{0.02}$ & 176.87$\pm{6.28}$&5 \\ 
E 3 & -7.09$\pm{1.73}$ & 3.38$\pm{1.92}$ & 99.0$\pm{99.0}$ & 10.88$\pm{1.28}$ & 4.49$\pm{1.56}$ & 0.44$\pm{0.14}$ & 0.47$\pm{0.14}$ & 0.08$\pm{0.03}$ & 0.73$\pm{0.21}$ & 24.59$\pm{2.04}$&8 \\ 
Pal 3 & 0.33$\pm{0.23}$ & 0.3$\pm{0.31}$ & 83.4$\pm{8.4}$ & 2053.2$\pm{462.51}$ & 94.05$\pm{3.22}$ & 0.91$\pm{0.02}$ & 0.63$\pm{0.04}$ & 0.5$\pm{0.01}$ & 0.0$\pm{0.01}$ & 234.07$\pm{10.37}$&8 \\ 
NGC 3201 & 5.28$\pm{0.32}$ & -0.98$\pm{0.33}$ & 494.0$\pm{0.2}$ & 21.34$\pm{0.99}$ & 9.0$\pm{0.08}$ & 0.41$\pm{0.02}$ & 3.17$\pm{0.2}$ & 0.5$\pm{0.01}$ & 0.01$\pm{0.01}$ & 83.46$\pm{2.35}$&5 \\ 
Pal 4* & -- & -- & -- & 140.18$\pm{25.72}$ & 46.43$\pm{23.97}$ & 0.53$\pm{0.18}$ & 0.84$\pm{0.06}$ & 0.5$\pm{0.01}$ & 0.67$\pm{0.27}$ & 283.04$\pm{12.73}$&-- \\ 
NGC 4147 & -2.08$\pm{0.48}$ & -3.07$\pm{0.46}$ & 183.2$\pm{0.7}$ & 29.31$\pm{2.12}$ & 7.02$\pm{2.05}$ & 0.62$\pm{0.08}$ & 1.09$\pm{0.06}$ & 0.41$\pm{0.03}$ & 0.65$\pm{0.07}$ & 95.97$\pm{5.03}$&3 \\ 
NGC 4372 & -6.49$\pm{0.33}$ & 3.71$\pm{0.32}$ & 72.3$\pm{1.2}$ & 7.41$\pm{0.06}$ & 2.99$\pm{0.11}$ & 0.43$\pm{0.01}$ & 4.59$\pm{0.26}$ & 0.43$\pm{0.01}$ & 0.98$\pm{0.01}$ & 77.82$\pm{2.31}$&5 \\ 
Rup 106 & -6.55$\pm{1.07}$ & 0.72$\pm{1.13}$ & -44.0$\pm{3.0}$ & 1946.97$\pm{290.22}$ & 18.47$\pm{0.69}$ & 0.98$\pm{0.01}$ & 1.16$\pm{0.08}$ & 0.5$\pm{0.01}$ & 0.0$\pm{0.01}$ & 96.01$\pm{4.8}$&8 \\ 
NGC 4590 & -3.76$\pm{0.66}$ & 1.79$\pm{0.62}$ & -94.7$\pm{0.2}$ & 36.7$\pm{5.23}$ & 9.66$\pm{0.31}$ & 0.58$\pm{0.04}$ & 2.94$\pm{0.19}$ & 0.5$\pm{0.01}$ & 0.02$\pm{0.01}$ & 88.54$\pm{3.36}$&2,3 \\ 
NGC 4833 & -8.11$\pm{0.35}$ & -0.96$\pm{9.34}$ & 200.2$\pm{1.2}$ & 8.57$\pm{0.44}$ & 0.98$\pm{0.54}$ & 0.8$\pm{0.1}$ & 7.11$\pm{0.69}$ & 0.38$\pm{0.06}$ & 0.82$\pm{0.06}$ & 85.12$\pm{3.55}$&5 \\ 
NGC 5024 & 0.5$\pm{1.0}$ & -0.1$\pm{1.0}$ & -62.9$\pm{0.3}$ & 55.37$\pm{17.46}$ & 16.85$\pm{0.93}$ & 0.54$\pm{0.11}$ & 9.79$\pm{0.67}$ & 0.53$\pm{0.01}$ & 0.04$\pm{0.03}$ & 199.02$\pm{8.8}$&3 \\ 
NGC 5053 & -1.89$\pm{1.23}$ & -0.39$\pm{1.95}$ & 44.0$\pm{0.4}$ & 19.84$\pm{1.9}$ & 8.86$\pm{3.42}$ & 0.42$\pm{0.12}$ & 1.77$\pm{0.11}$ & 0.46$\pm{0.02}$ & 0.85$\pm{0.14}$ & 103.63$\pm{4.91}$&8 \\ 
NGC 5139 & -5.08$\pm{0.35}$ & -3.57$\pm{0.34}$ & 232.1$\pm{0.1}$ & 6.65$\pm{0.03}$ & 0.97$\pm{0.04}$ & 0.75$\pm{0.01}$ & 41.19$\pm{2.54}$ & 0.53$\pm{0.01}$ & 0.99$\pm{0.01}$ & 161.71$\pm{3.66}$&2,3 \\ 
NGC 5272 & -1.1$\pm{0.51}$ & -2.3$\pm{0.54}$ & -147.6$\pm{0.2}$ & 14.75$\pm{0.63}$ & 5.03$\pm{0.95}$ & 0.49$\pm{0.07}$ & 11.61$\pm{0.75}$ & 0.52$\pm{0.01}$ & 0.74$\pm{0.05}$ & 158.64$\pm{5.67}$&3 \\ 
NGC 5286 & -5.09$\pm{0.92}$ & -1.2$\pm{0.88}$ & 57.4$\pm{1.5}$ & 8.99$\pm{0.28}$ & 4.76$\pm{0.96}$ & 0.3$\pm{0.09}$ & 10.28$\pm{0.63}$ & 0.51$\pm{0.01}$ & 0.98$\pm{0.02}$ & 121.62$\pm{5.16}$&8 \\ 
AM 4* & -- & -- & -- & 36.94$\pm{7.98}$ & 16.91$\pm{5.48}$ & 0.41$\pm{0.12}$ & 0.06$\pm{0.02}$ & 0.08$\pm{0.03}$ & 0.49$\pm{0.3}$ & 25.0$\pm{2.11}$&-- \\ 
NGC 5466 & -4.65$\pm{0.82}$ & 0.8$\pm{0.82}$ & 110.7$\pm{0.2}$ & 80.14$\pm{25.9}$ & 6.77$\pm{1.08}$ & 0.85$\pm{0.02}$ & 2.1$\pm{0.12}$ & 0.47$\pm{0.01}$ & 0.13$\pm{0.06}$ & 105.03$\pm{4.52}$&3 \\ 
NGC 5634 & -5.3$\pm{3.02}$ & -0.65$\pm{2.12}$ & -45.1$\pm{6.6}$ & 2470.22$\pm{697.16}$ & 21.01$\pm{0.72}$ & 0.98$\pm{0.01}$ & 3.87$\pm{0.23}$ & 0.52$\pm{0.01}$ & 0.0$\pm{0.01}$ & 158.46$\pm{6.69}$&8 \\ 
NGC 5694* & -- & -- & -- & 34.32$\pm{4.79}$ & 11.46$\pm{5.85}$ & 0.53$\pm{0.18}$ & 4.47$\pm{0.29}$ & 0.51$\pm{0.01}$ & 0.76$\pm{0.21}$ & 203.68$\pm{9.94}$&-- \\ 
IC 4499* & -- & -- & -- & 18.4$\pm{2.57}$ & 6.38$\pm{3.28}$ & 0.49$\pm{0.18}$ & 2.94$\pm{0.23}$ & 0.47$\pm{0.02}$ & 0.75$\pm{0.21}$ & 112.86$\pm{6.44}$&8 \\ 
NGC 5824* & -- & -- & -- & 30.21$\pm{4.19}$ & 9.82$\pm{5.21}$ & 0.53$\pm{0.18}$ & 11.26$\pm{0.73}$ & 0.53$\pm{0.01}$ & 0.77$\pm{0.2}$ & 258.04$\pm{12.75}$&-- \\ 
Pal 5 & -2.39$\pm{0.17}$ & -2.36$\pm{0.15}$ & -58.7$\pm{0.2}$ & 18.9$\pm{0.66}$ & 8.16$\pm{1.42}$ & 0.4$\pm{0.07}$ & 0.49$\pm{0.02}$ & 0.31$\pm{0.03}$ & 0.96$\pm{0.01}$ & 60.89$\pm{3.31}$&9 \\ 
NGC 5897 & -4.93$\pm{0.86}$ & -2.33$\pm{0.84}$ & 101.5$\pm{1.0}$ & 9.26$\pm{0.79}$ & 1.87$\pm{0.51}$ & 0.67$\pm{0.05}$ & 3.18$\pm{0.17}$ & 0.33$\pm{0.04}$ & 0.75$\pm{0.07}$ & 62.99$\pm{4.14}$&2,3 \\ 
NGC 5904 & 5.07$\pm{0.68}$ & 10.7$\pm{0.56}$ & 53.2$\pm{0.4}$ & 57.57$\pm{10.7}$ & 3.43$\pm{0.13}$ & 0.89$\pm{0.02}$ & 10.93$\pm{0.67}$ & 0.52$\pm{0.01}$ & 0.05$\pm{0.01}$ & 100.52$\pm{2.97}$&11 \\ 
NGC 5927 & -5.72$\pm{0.39}$ & -2.61$\pm{0.4}$ & -107.5$\pm{0.9}$ & 5.1$\pm{0.16}$ & 4.47$\pm{0.12}$ & 0.07$\pm{0.02}$ & 4.57$\pm{0.27}$ & 0.46$\pm{0.01}$ & 0.35$\pm{0.09}$ & 58.79$\pm{1.77}$&5 \\ 
NGC 5946 & -4.39$\pm{0.3}$ & -6.01$\pm{0.26}$ & 128.4$\pm{1.8}$ & 6.83$\pm{0.41}$ & 1.46$\pm{0.32}$ & 0.65$\pm{0.05}$ & 3.28$\pm{0.06}$ & 0.29$\pm{0.04}$ & 0.81$\pm{0.04}$ & 52.05$\pm{3.31}$&8 \\ 
NGC 5986 & -3.81$\pm{0.45}$ & -2.99$\pm{0.37}$ & 88.9$\pm{3.7}$ & 5.09$\pm{0.26}$ & 0.23$\pm{0.06}$ & 0.91$\pm{0.02}$ & 11.65$\pm{1.03}$ & 0.25$\pm{0.04}$ & 0.96$\pm{0.01}$ & 66.05$\pm{4.83}$&5 \\ 
Pal 14* & -- & -- & -- & 88.01$\pm{15.42}$ & 33.71$\pm{15.68}$ & 0.48$\pm{0.16}$ & 0.29$\pm{0.02}$ & 0.46$\pm{0.02}$ & 0.64$\pm{0.28}$ & 141.39$\pm{7.14}$&-- \\ 
Lynga 7* & -- & -- & -- & 4.99$\pm{0.44}$ & 1.48$\pm{0.64}$ & 0.55$\pm{0.15}$ & 2.35$\pm{0.35}$ & 0.2$\pm{0.05}$ & 0.83$\pm{0.13}$ & 34.39$\pm{1.79}$&-- \\ 
NGC 6093 & -3.31$\pm{0.58}$ & -7.2$\pm{0.67}$ & 8.1$\pm{1.5}$ & 3.89$\pm{0.21}$ & 1.88$\pm{0.51}$ & 0.35$\pm{0.11}$ & 7.0$\pm{0.33}$ & 0.42$\pm{0.02}$ & 0.9$\pm{0.05}$ & 55.84$\pm{3.61}$&2,3 \\ 
NGC 6121 & -12.5$\pm{0.36}$ & -19.93$\pm{0.49}$ & 70.7$\pm{0.2}$ & 6.34$\pm{0.06}$ & 0.24$\pm{0.08}$ & 0.93$\pm{0.02}$ & 6.46$\pm{1.26}$ & 0.11$\pm{0.04}$ & 0.97$\pm{0.01}$ & 50.01$\pm{2.46}$&2,3 \\ 

\hline
\end{tabular}
\caption{Summary of the orbital parameters obtained from the sampling of
positions and velocities for the clusters in our sample. The reference for the
PM measurment is given in the last column. \comm{Clusters in the noPM are marked
with * after their name.} The reference IDs are the following: (1)
\citet{Dinescu97}; (2) \citet{Dinescu99}; (3) \citet{Dinescu99b}; (4)
\citet{Dinescu03}; (5) \citet{Dinescu07}; (6) \citet{Dinescu10}; (7)
\citet{Dinescu13}; (8) \citet{Dambis06}; (9) \citet{Kuepper15}; (10)
\citet{Rossi15}; and (11) \citet{Scholtz96}.}
\label{summary}
\end{table}
\end{landscape}

\begin{landscape}
\begin{table}
\footnotesize
\begin{tabular}{lccccccccccc}
\hline
Name & $\mu_{\alpha}$ & $\mu_{\delta}$ & V$_{\rm los}$ & R$_{apo}$ & R$_{peri}$ & ecc & M$_{i}$ & $\mu$ & $\phi$ & $r_J$ & Refs \\\\
 & mas/yr & mas/yr & km$\,$ s$^{-1}$ & kpc & kpc & & $10^5$ M$_{\odot}$ & & & pc & \\
\hline
\hline
NGC 6101 & -1.77$\pm{1.49}$ & -4.65$\pm{1.41}$ & 361.4$\pm{1.7}$ & 38.09$\pm{13.11}$ & 3.77$\pm{1.21}$ & 0.84$\pm{0.04}$ & 2.2$\pm{0.16}$ & 0.41$\pm{0.04}$ & 0.21$\pm{0.09}$ & 78.02$\pm{4.56}$&8 \\ 
NGC 6144 & -3.06$\pm{0.64}$ & -5.11$\pm{0.72}$ & 193.8$\pm{0.6}$ & 3.23$\pm{0.16}$ & 1.77$\pm{0.09}$ & 0.29$\pm{0.04}$ & 2.75$\pm{0.09}$ & 0.21$\pm{0.02}$ & 0.64$\pm{0.02}$ & 26.68$\pm{1.37}$&2,3 \\ 
NGC 6139 & -3.65$\pm{0.38}$ & -8.34$\pm{0.36}$ & 6.7$\pm{6.0}$ & 4.28$\pm{0.4}$ & 2.24$\pm{0.43}$ & 0.33$\pm{0.04}$ & 7.65$\pm{0.36}$ & 0.45$\pm{0.02}$ & 0.61$\pm{0.12}$ & 56.95$\pm{3.9}$&8 \\ 
Terzan 3 & -2.95$\pm{1.41}$ & -6.35$\pm{1.35}$ & -136.3$\pm{0.7}$ & 3.0$\pm{0.22}$ & 2.1$\pm{0.16}$ & 0.18$\pm{0.06}$ & 1.41$\pm{0.04}$ & 0.05$\pm{0.02}$ & 0.52$\pm{0.1}$ & 12.8$\pm{1.66}$&8 \\ 
NGC 6171 & -0.7$\pm{0.9}$ & -3.1$\pm{1.0}$ & -34.1$\pm{0.3}$ & 3.6$\pm{0.09}$ & 2.26$\pm{0.32}$ & 0.23$\pm{0.07}$ & 3.01$\pm{0.18}$ & 0.3$\pm{0.02}$ & 0.96$\pm{0.03}$ & 36.49$\pm{0.76}$&3 \\ 
NGC 6205 & -0.9$\pm{0.71}$ & 5.5$\pm{1.12}$ & -244.2$\pm{0.2}$ & 25.39$\pm{5.23}$ & 5.65$\pm{0.31}$ & 0.64$\pm{0.06}$ & 8.6$\pm{0.53}$ & 0.52$\pm{0.01}$ & 0.15$\pm{0.04}$ & 113.26$\pm{3.51}$&3 \\ 
NGC 6229* & -- & -- & -- & 34.43$\pm{4.26}$ & 11.81$\pm{6.09}$ & 0.51$\pm{0.18}$ & 5.5$\pm{0.37}$ & 0.52$\pm{0.01}$ & 0.78$\pm{0.19}$ & 222.63$\pm{10.08}$&-- \\ 
NGC 6218 & 1.3$\pm{0.58}$ & -7.83$\pm{0.62}$ & -41.4$\pm{0.2}$ & 5.39$\pm{0.12}$ & 2.47$\pm{0.24}$ & 0.37$\pm{0.04}$ & 3.33$\pm{0.2}$ & 0.34$\pm{0.01}$ & 0.81$\pm{0.04}$ & 48.74$\pm{0.82}$&3 \\ 
NGC 6235* & -- & -- & -- & 5.02$\pm{0.91}$ & 1.89$\pm{0.7}$ & 0.46$\pm{0.12}$ & 1.89$\pm{0.27}$ & 0.18$\pm{0.04}$ & 0.66$\pm{0.24}$ & 28.76$\pm{2.67}$&-- \\ 
NGC 6254 & -6.0$\pm{1.0}$ & -3.3$\pm{1.0}$ & 75.2$\pm{0.7}$ & 5.02$\pm{0.1}$ & 3.13$\pm{0.33}$ & 0.23$\pm{0.05}$ & 3.64$\pm{0.22}$ & 0.39$\pm{0.01}$ & 0.91$\pm{0.01}$ & 52.97$\pm{0.69}$&3 \\ 
NGC 6256* & -- & -- & -- & 3.5$\pm{0.58}$ & 1.1$\pm{0.45}$ & 0.54$\pm{0.14}$ & 3.52$\pm{0.37}$ & 0.23$\pm{0.04}$ & 0.75$\pm{0.18}$ & 30.76$\pm{3.03}$&-- \\ 
Pal 15* & -- & -- & -- & 46.05$\pm{7.23}$ & 17.87$\pm{7.89}$ & 0.46$\pm{0.15}$ & 0.56$\pm{0.04}$ & 0.45$\pm{0.03}$ & 0.68$\pm{0.26}$ & 115.61$\pm{6.11}$&-- \\ 
NGC 6266 & -3.5$\pm{0.37}$ & -0.82$\pm{0.37}$ & -70.1$\pm{1.4}$ & 2.24$\pm{0.17}$ & 1.77$\pm{0.14}$ & 0.12$\pm{0.02}$ & 15.83$\pm{1.02}$ & 0.48$\pm{0.01}$ & 0.47$\pm{0.11}$ & 51.14$\pm{1.47}$&4 \\ 
NGC 6273 & -2.86$\pm{0.49}$ & -0.45$\pm{0.51}$ & 135.0$\pm{4.1}$ & 2.85$\pm{0.2}$ & 0.72$\pm{0.21}$ & 0.59$\pm{0.09}$ & 15.65$\pm{0.46}$ & 0.44$\pm{0.02}$ & 0.41$\pm{0.04}$ & 42.2$\pm{2.99}$&6 \\ 
NGC 6284 & -3.66$\pm{0.64}$ & -5.39$\pm{0.83}$ & 27.5$\pm{1.7}$ & 8.23$\pm{1.23}$ & 5.97$\pm{1.08}$ & 0.21$\pm{0.06}$ & 5.1$\pm{0.27}$ & 0.48$\pm{0.01}$ & 0.56$\pm{0.36}$ & 82.39$\pm{5.57}$&6 \\ 
NGC 6287 & -3.68$\pm{0.88}$ & -3.54$\pm{0.69}$ & -288.7$\pm{3.5}$ & 5.3$\pm{0.37}$ & 0.24$\pm{0.07}$ & 0.91$\pm{0.02}$ & 7.04$\pm{1.3}$ & 0.12$\pm{0.04}$ & 0.37$\pm{0.02}$ & 25.46$\pm{2.83}$&6 \\ 
NGC 6293 & 0.26$\pm{0.85}$ & -5.14$\pm{0.71}$ & -146.2$\pm{1.7}$ & 3.28$\pm{0.35}$ & 0.31$\pm{0.09}$ & 0.82$\pm{0.04}$ & 7.34$\pm{0.84}$ & 0.19$\pm{0.04}$ & 0.53$\pm{0.06}$ & 27.99$\pm{3.45}$&6 \\ 
NGC 6304 & -2.59$\pm{0.29}$ & -1.56$\pm{0.29}$ & -107.3$\pm{3.6}$ & 3.09$\pm{0.18}$ & 1.9$\pm{0.1}$ & 0.24$\pm{0.01}$ & 3.41$\pm{0.2}$ & 0.31$\pm{0.01}$ & 0.56$\pm{0.04}$ & 31.43$\pm{0.68}$&4 \\ 
NGC 6316 & -2.42$\pm{0.63}$ & -1.71$\pm{0.56}$ & 71.4$\pm{8.9}$ & 2.59$\pm{0.39}$ & 1.13$\pm{0.22}$ & 0.38$\pm{0.1}$ & 8.09$\pm{0.41}$ & 0.39$\pm{0.02}$ & 0.83$\pm{0.05}$ & 42.58$\pm{4.85}$&4 \\ 
NGC 6341 & -3.3$\pm{0.55}$ & -0.33$\pm{0.7}$ & -120.0$\pm{0.1}$ & 10.58$\pm{0.29}$ & 0.82$\pm{0.25}$ & 0.86$\pm{0.04}$ & 7.46$\pm{0.53}$ & 0.36$\pm{0.03}$ & 0.92$\pm{0.02}$ & 103.5$\pm{3.9}$&3 \\ 
NGC 6325 & -3.91$\pm{0.85}$ & -5.4$\pm{0.74}$ & 29.8$\pm{1.8}$ & 1.24$\pm{0.04}$ & 0.24$\pm{0.05}$ & 0.68$\pm{0.06}$ & 7.32$\pm{1.12}$ & 0.08$\pm{0.02}$ & 0.99$\pm{0.01}$ & 15.66$\pm{1.41}$&8 \\ 
NGC 6333 & -0.57$\pm{0.57}$ & -3.7$\pm{0.5}$ & 229.1$\pm{7.0}$ & 4.76$\pm{0.23}$ & 1.04$\pm{0.08}$ & 0.64$\pm{0.02}$ & 5.9$\pm{0.36}$ & 0.35$\pm{0.01}$ & 0.2$\pm{0.02}$ & 30.82$\pm{0.47}$&6 \\ 
NGC 6342 & -2.77$\pm{0.71}$ & -5.84$\pm{0.65}$ & 115.7$\pm{1.4}$ & 2.0$\pm{0.08}$ & 0.93$\pm{0.04}$ & 0.36$\pm{0.02}$ & 2.92$\pm{0.03}$ & 0.11$\pm{0.02}$ & 0.68$\pm{0.03}$ & 15.95$\pm{1.24}$&6 \\ 
NGC 6356 & -3.14$\pm{0.68}$ & -3.65$\pm{0.53}$ & 27.0$\pm{4.3}$ & 7.56$\pm{0.46}$ & 2.18$\pm{0.8}$ & 0.55$\pm{0.12}$ & 8.65$\pm{0.4}$ & 0.47$\pm{0.02}$ & 0.92$\pm{0.02}$ & 96.26$\pm{6.21}$&6 \\ 
NGC 6355* & -- & -- & -- & 2.12$\pm{0.62}$ & 0.57$\pm{0.19}$ & 0.55$\pm{0.12}$ & 7.33$\pm{0.4}$ & 0.29$\pm{0.04}$ & 0.47$\pm{0.22}$ & 25.03$\pm{3.69}$&-- \\ 
NGC 6352* & -- & -- & -- & 4.14$\pm{0.44}$ & 1.31$\pm{0.57}$ & 0.52$\pm{0.16}$ & 2.22$\pm{0.36}$ & 0.18$\pm{0.04}$ & 0.79$\pm{0.15}$ & 28.4$\pm{0.9}$&-- \\ 
IC 1257 & -4.81$\pm{0.64}$ & -3.44$\pm{0.64}$ & -140.2$\pm{2.1}$ & 1791.9$\pm{426.29}$ & 17.51$\pm{0.76}$ & 0.98$\pm{0.01}$ & 0.98$\pm{0.06}$ & 0.49$\pm{0.01}$ & 0.0$\pm{0.01}$ & 87.81$\pm{4.46}$&8 \\ 
Terzan 2 & -0.94$\pm{0.3}$ & 0.15$\pm{0.42}$ & 109.0$\pm{15.0}$ & 2.59$\pm{0.24}$ & 0.45$\pm{0.15}$ & 0.7$\pm{0.07}$ & 3.68$\pm{0.9}$ & 0.05$\pm{0.02}$ & 0.26$\pm{0.02}$ & 9.6$\pm{1.46}$&10 \\ 
NGC 6366 & -3.9$\pm{0.57}$ & -6.13$\pm{0.52}$ & -122.2$\pm{0.5}$ & 5.69$\pm{0.11}$ & 1.15$\pm{0.12}$ & 0.66$\pm{0.02}$ & 1.81$\pm{0.12}$ & 0.09$\pm{0.01}$ & 0.92$\pm{0.01}$ & 27.73$\pm{1.41}$&8 \\ 
Terzan 4 & 3.5$\pm{0.69}$ & 0.35$\pm{0.58}$ & -50.0$\pm{2.9}$ & 4.17$\pm{0.4}$ & 1.17$\pm{0.21}$ & 0.56$\pm{0.06}$ & 1.71$\pm{0.21}$ & 0.03$\pm{0.02}$ & 0.03$\pm{0.01}$ & 7.53$\pm{1.35}$&10 \\ 
NGC 6362 & -3.09$\pm{0.46}$ & -3.83$\pm{0.46}$ & -13.1$\pm{0.6}$ & 5.51$\pm{0.11}$ & 2.22$\pm{0.09}$ & 0.43$\pm{0.02}$ & 2.61$\pm{0.12}$ & 0.28$\pm{0.01}$ & 0.92$\pm{0.03}$ & 44.76$\pm{1.5}$&1,3 \\ 
Liller 1 & -4.6$\pm{0.83}$ & -4.12$\pm{0.79}$ & 52.0$\pm{15.0}$ & 0.9$\pm{0.07}$ & 0.17$\pm{0.04}$ & 0.69$\pm{0.07}$ & 10.24$\pm{1.84}$ & 0.07$\pm{0.03}$ & 0.88$\pm{0.05}$ & 12.86$\pm{1.43}$&8 \\ 
NGC 6380* & -- & -- & -- & 3.82$\pm{0.62}$ & 1.17$\pm{0.5}$ & 0.53$\pm{0.15}$ & 4.3$\pm{0.37}$ & 0.29$\pm{0.05}$ & 0.76$\pm{0.18}$ & 37.52$\pm{3.77}$&-- \\ 
Terzan 1 & 0.51$\pm{0.31}$ & -0.93$\pm{0.29}$ & 114.0$\pm{14.0}$ & 3.13$\pm{0.24}$ & 1.09$\pm{0.15}$ & 0.48$\pm{0.04}$ & 1.88$\pm{0.18}$ & 0.03$\pm{0.02}$ & 0.27$\pm{0.02}$ & 8.91$\pm{1.67}$&10 \\ 
Ton 2 & -2.97$\pm{0.81}$ & -3.89$\pm{1.02}$ & -184.4$\pm{2.2}$ & 2.28$\pm{0.14}$ & 1.28$\pm{0.09}$ & 0.28$\pm{0.05}$ & 2.4$\pm{0.09}$ & 0.11$\pm{0.02}$ & 0.16$\pm{0.06}$ & 13.41$\pm{0.89}$&8 \\ 
NGC 6388 & -1.9$\pm{0.45}$ & -3.83$\pm{0.51}$ & 80.1$\pm{0.8}$ & 3.11$\pm{0.15}$ & 0.72$\pm{0.08}$ & 0.62$\pm{0.03}$ & 19.67$\pm{1.04}$ & 0.47$\pm{0.01}$ & 0.93$\pm{0.03}$ & 70.95$\pm{4.64}$&6 \\ 
NGC 6402 & -0.13$\pm{0.51}$ & -5.42$\pm{0.51}$ & -66.1$\pm{1.8}$ & 4.26$\pm{0.17}$ & 1.14$\pm{0.22}$ & 0.58$\pm{0.05}$ & 14.85$\pm{0.88}$ & 0.47$\pm{0.01}$ & 0.93$\pm{0.03}$ & 79.77$\pm{3.85}$&8 \\ 
NGC 6401* & -- & -- & -- & 3.11$\pm{0.56}$ & 0.99$\pm{0.4}$ & 0.51$\pm{0.14}$ & 5.9$\pm{0.41}$ & 0.33$\pm{0.04}$ & 0.7$\pm{0.21}$ & 37.14$\pm{4.39}$&-- \\ 
NGC 6397 & 3.69$\pm{0.29}$ & -14.88$\pm{0.26}$ & 18.8$\pm{0.1}$ & 6.43$\pm{0.05}$ & 3.12$\pm{0.1}$ & 0.35$\pm{0.01}$ & 1.97$\pm{0.1}$ & 0.28$\pm{0.01}$ & 0.95$\pm{0.01}$ & 45.96$\pm{0.88}$&3,7 \\ 
Pal 6 & 2.95$\pm{0.41}$ & 1.24$\pm{0.19}$ & 181.0$\pm{2.8}$ & 8.45$\pm{0.46}$ & 2.16$\pm{0.13}$ & 0.59$\pm{0.01}$ & 2.26$\pm{0.14}$ & 0.29$\pm{0.01}$ & 0.06$\pm{0.01}$ & 26.16$\pm{0.76}$&10 \\ 
NGC 6426 & -4.05$\pm{1.47}$ & -6.63$\pm{0.65}$ & -162.0$\pm{23.0}$ & 1819.11$\pm{415.72}$ & 13.3$\pm{0.65}$ & 0.99$\pm{0.01}$ & 1.55$\pm{0.1}$ & 0.5$\pm{0.01}$ & 0.0$\pm{0.01}$ & 88.74$\pm{4.72}$&8 \\ 
Djorg 1 & 3.14$\pm{1.53}$ & -6.76$\pm{0.68}$ & -362.4$\pm{3.6}$ & 208.24$\pm{86.31}$ & 3.7$\pm{0.38}$ & 0.97$\pm{0.01}$ & 2.18$\pm{0.11}$ & 0.42$\pm{0.02}$ & 0.01$\pm{0.01}$ & 49.07$\pm{4.05}$&8 \\ 
Terzan 5 & -4.05$\pm{0.87}$ & -2.65$\pm{0.78}$ & -93.0$\pm{2.0}$ & 1.91$\pm{0.19}$ & 0.49$\pm{0.11}$ & 0.59$\pm{0.05}$ & 5.29$\pm{0.57}$ & 0.18$\pm{0.02}$ & 0.74$\pm{0.05}$ & 21.49$\pm{1.64}$&8 \\ 
NGC 6440* & -- & -- & -- & 1.94$\pm{0.4}$ & 0.57$\pm{0.23}$ & 0.57$\pm{0.12}$ & 12.03$\pm{0.84}$ & 0.38$\pm{0.04}$ & 0.53$\pm{0.21}$ & 32.02$\pm{1.6}$&-- \\ 
NGC 6441 & -2.86$\pm{0.45}$ & -3.45$\pm{0.76}$ & 16.5$\pm{1.0}$ & 3.66$\pm{0.36}$ & 0.6$\pm{0.08}$ & 0.71$\pm{0.05}$ & 23.99$\pm{1.47}$ & 0.48$\pm{0.01}$ & 0.98$\pm{0.02}$ & 86.88$\pm{7.88}$&6 \\ 

\hline
\end{tabular}
\contcaption{}
\label{summary}
\end{table}
\end{landscape}

\begin{landscape}
\begin{table}
\footnotesize
\begin{tabular}{lccccccccccc}
\hline
Name & $\mu_{\alpha}$ & $\mu_{\delta}$ & V$_{\rm los}$ & R$_{apo}$ & R$_{peri}$ & ecc & M$_{i}$ & $\mu$ & $\phi$ & $r_J$ & Refs \\\\
 & mas/yr & mas/yr & km$\,$ s$^{-1}$ & kpc & kpc & & $10^5$ M$_{\odot}$ & & & pc & \\
\hline
\hline
Terzan 6* & -- & -- & -- & 2.13$\pm{0.43}$ & 0.74$\pm{0.28}$ & 0.48$\pm{0.15}$ & 5.2$\pm{0.6}$ & 0.24$\pm{0.04}$ & 0.57$\pm{0.26}$ & 23.38$\pm{1.8}$&-- \\ 
NGC 6453 & -1.09$\pm{0.78}$ & -2.14$\pm{0.84}$ & -83.7$\pm{8.3}$ & 4.05$\pm{0.41}$ & 1.11$\pm{0.26}$ & 0.56$\pm{0.07}$ & 3.62$\pm{0.2}$ & 0.25$\pm{0.04}$ & 0.78$\pm{0.05}$ & 35.88$\pm{3.95}$&8 \\ 
UKS 1 & -3.48$\pm{1.57}$ & -4.34$\pm{1.2}$ & 57.0$\pm{6.0}$ & 1.0$\pm{0.11}$ & 0.48$\pm{0.11}$ & 0.32$\pm{0.1}$ & 5.04$\pm{0.64}$ & 0.11$\pm{0.02}$ & 0.86$\pm{0.08}$ & 12.31$\pm{1.06}$&8 \\ 
NGC 6496 & -2.39$\pm{1.0}$ & -6.64$\pm{1.02}$ & -112.7$\pm{5.7}$ & 4.76$\pm{0.52}$ & 2.97$\pm{0.53}$ & 0.24$\pm{0.05}$ & 2.88$\pm{0.13}$ & 0.37$\pm{0.03}$ & 0.57$\pm{0.17}$ & 41.48$\pm{3.45}$&8 \\ 
Terzan 9 & 0.0$\pm{0.38}$ & -3.07$\pm{0.49}$ & 59.0$\pm{10.0}$ & 1.54$\pm{0.18}$ & 1.23$\pm{0.15}$ & 0.11$\pm{0.03}$ & 2.08$\pm{0.2}$ & 0.03$\pm{0.02}$ & 0.23$\pm{0.15}$ & 7.25$\pm{1.66}$&10 \\ 
Djorg 2 & 0.99$\pm{1.15}$ & -1.9$\pm{1.41}$ & 99.0$\pm{99.0}$ & 2.65$\pm{0.46}$ & 0.91$\pm{0.35}$ & 0.49$\pm{0.14}$ & 3.58$\pm{0.48}$ & 0.19$\pm{0.04}$ & 0.67$\pm{0.21}$ & 22.87$\pm{1.26}$&8 \\ 
NGC 6517 & -4.86$\pm{0.93}$ & -4.99$\pm{0.86}$ & -39.6$\pm{8.0}$ & 4.17$\pm{0.26}$ & 1.47$\pm{0.38}$ & 0.47$\pm{0.08}$ & 7.17$\pm{0.34}$ & 0.41$\pm{0.03}$ & 0.94$\pm{0.04}$ & 59.14$\pm{4.28}$&8 \\ 
Terzan 10 & 1.88$\pm{0.19}$ & -3.41$\pm{0.22}$ & 99.0$\pm{99.0}$ & 3.13$\pm{0.45}$ & 1.06$\pm{0.44}$ & 0.5$\pm{0.15}$ & 2.57$\pm{0.45}$ & 0.13$\pm{0.03}$ & 0.73$\pm{0.19}$ & 20.99$\pm{1.24}$&10 \\ 
NGC 6522 & 6.08$\pm{0.2}$ & -1.83$\pm{0.2}$ & -21.1$\pm{3.4}$ & 4.3$\pm{0.18}$ & 0.66$\pm{0.2}$ & 0.73$\pm{0.06}$ & 5.17$\pm{0.57}$ & 0.26$\pm{0.03}$ & 0.05$\pm{0.01}$ & 16.19$\pm{1.82}$&4 \\ 
NGC 6535 & 3.3$\pm{0.83}$ & -6.85$\pm{0.83}$ & -215.1$\pm{0.5}$ & 4.86$\pm{0.22}$ & 2.97$\pm{0.48}$ & 0.25$\pm{0.07}$ & 1.03$\pm{0.11}$ & 0.06$\pm{0.01}$ & 0.56$\pm{0.15}$ & 16.84$\pm{1.08}$&8 \\ 
NGC 6528 & -0.35$\pm{0.23}$ & 0.27$\pm{0.26}$ & 206.6$\pm{1.4}$ & 3.79$\pm{0.19}$ & 0.55$\pm{0.03}$ & 0.73$\pm{0.02}$ & 3.6$\pm{0.13}$ & 0.11$\pm{0.02}$ & 0.06$\pm{0.02}$ & 9.7$\pm{0.98}$&4 \\ 
NGC 6539 & -6.24$\pm{0.55}$ & -1.23$\pm{0.64}$ & 31.0$\pm{1.7}$ & 4.24$\pm{0.28}$ & 1.47$\pm{0.13}$ & 0.48$\pm{0.05}$ & 7.43$\pm{0.45}$ & 0.41$\pm{0.01}$ & 0.6$\pm{0.06}$ & 50.76$\pm{1.29}$&8 \\ 
NGC 6540 & 0.07$\pm{0.4}$ & 1.9$\pm{0.57}$ & -17.7$\pm{1.4}$ & 4.9$\pm{0.32}$ & 2.97$\pm{0.17}$ & 0.24$\pm{0.03}$ & 1.69$\pm{0.1}$ & 0.23$\pm{0.02}$ & 0.03$\pm{0.01}$ & 25.22$\pm{0.64}$&10 \\ 
NGC 6544* & -- & -- & -- & 5.88$\pm{0.45}$ & 1.78$\pm{0.71}$ & 0.54$\pm{0.14}$ & 2.69$\pm{0.3}$ & 0.27$\pm{0.05}$ & 0.85$\pm{0.12}$ & 45.13$\pm{1.51}$&-- \\ 
NGC 6541* & -- & -- & -- & 2.92$\pm{0.49}$ & 0.96$\pm{0.42}$ & 0.53$\pm{0.15}$ & 9.62$\pm{0.69}$ & 0.39$\pm{0.03}$ & 0.67$\pm{0.22}$ & 43.9$\pm{0.7}$&-- \\ 
NGC 6553 & 2.5$\pm{0.07}$ & 5.35$\pm{0.08}$ & -3.2$\pm{1.5}$ & 11.09$\pm{0.2}$ & 2.33$\pm{0.18}$ & 0.65$\pm{0.02}$ & 4.56$\pm{0.27}$ & 0.42$\pm{0.01}$ & 0.01$\pm{0.01}$ & 36.93$\pm{0.99}$&4 \\ 
NGC 6558 & -0.12$\pm{0.55}$ & 0.47$\pm{0.6}$ & -197.2$\pm{1.5}$ & 4.25$\pm{0.33}$ & 1.05$\pm{0.09}$ & 0.61$\pm{0.02}$ & 2.54$\pm{0.15}$ & 0.14$\pm{0.02}$ & 0.05$\pm{0.02}$ & 13.17$\pm{0.97}$&10 \\ 
IC 1276 & -17.13$\pm{0.6}$ & -13.4$\pm{0.67}$ & 155.7$\pm{1.3}$ & 28.03$\pm{4.72}$ & 0.72$\pm{0.17}$ & 0.95$\pm{0.01}$ & 3.06$\pm{0.32}$ & 0.15$\pm{0.03}$ & 0.12$\pm{0.03}$ & 31.37$\pm{1.47}$&8 \\ 
Terzan 12* & -- & -- & -- & 4.23$\pm{0.47}$ & 1.41$\pm{0.58}$ & 0.5$\pm{0.15}$ & 1.5$\pm{0.41}$ & 0.03$\pm{0.02}$ & 0.77$\pm{0.18}$ & 13.79$\pm{2.53}$&-- \\ 
NGC 6569* & -- & -- & -- & 3.55$\pm{0.63}$ & 1.07$\pm{0.5}$ & 0.53$\pm{0.17}$ & 7.76$\pm{0.51}$ & 0.39$\pm{0.04}$ & 0.7$\pm{0.2}$ & 46.88$\pm{5.19}$&-- \\ 
NGC 6584 & -0.22$\pm{0.62}$ & -5.79$\pm{0.67}$ & 222.9$\pm{15.0}$ & 12.86$\pm{1.69}$ & 1.19$\pm{0.44}$ & 0.83$\pm{0.04}$ & 4.74$\pm{0.24}$ & 0.35$\pm{0.05}$ & 0.48$\pm{0.07}$ & 69.47$\pm{5.53}$&1,3 \\ 
NGC 6624* & -- & -- & -- & 1.95$\pm{0.49}$ & 0.5$\pm{0.12}$ & 0.61$\pm{0.1}$ & 5.32$\pm{0.55}$ & 0.21$\pm{0.03}$ & 0.52$\pm{0.2}$ & 19.66$\pm{0.62}$&-- \\ 
NGC 6626 & 0.63$\pm{0.67}$ & -8.46$\pm{0.67}$ & 17.0$\pm{1.0}$ & 3.17$\pm{0.18}$ & 0.78$\pm{0.15}$ & 0.6$\pm{0.06}$ & 7.21$\pm{0.5}$ & 0.34$\pm{0.02}$ & 0.95$\pm{0.04}$ & 46.4$\pm{0.85}$&3,7 \\ 
NGC 6638 & -0.56$\pm{1.48}$ & -4.06$\pm{1.38}$ & 18.1$\pm{3.9}$ & 2.52$\pm{0.42}$ & 0.32$\pm{0.1}$ & 0.76$\pm{0.07}$ & 5.72$\pm{1.08}$ & 0.12$\pm{0.04}$ & 0.79$\pm{0.09}$ & 22.6$\pm{2.9}$&8 \\ 
NGC 6637 & -3.51$\pm{1.29}$ & -2.4$\pm{1.2}$ & 39.9$\pm{2.8}$ & 2.05$\pm{0.24}$ & 0.67$\pm{0.12}$ & 0.49$\pm{0.07}$ & 5.19$\pm{0.17}$ & 0.26$\pm{0.03}$ & 0.71$\pm{0.12}$ & 24.93$\pm{2.14}$&8 \\ 
NGC 6642 & 0.52$\pm{1.08}$ & -3.72$\pm{1.2}$ & -57.2$\pm{5.4}$ & 2.49$\pm{0.19}$ & 0.25$\pm{0.06}$ & 0.81$\pm{0.04}$ & 6.13$\pm{1.06}$ & 0.07$\pm{0.02}$ & 0.66$\pm{0.06}$ & 17.44$\pm{1.93}$&8 \\ 
NGC 6652 & 4.75$\pm{0.07}$ & -4.45$\pm{0.1}$ & -111.7$\pm{5.8}$ & 8.1$\pm{0.7}$ & 2.17$\pm{0.23}$ & 0.58$\pm{0.01}$ & 2.1$\pm{0.01}$ & 0.28$\pm{0.02}$ & 0.08$\pm{0.02}$ & 26.3$\pm{1.91}$&10 \\ 
NGC 6656 & 7.37$\pm{0.5}$ & -3.95$\pm{0.42}$ & -146.3$\pm{0.2}$ & 9.14$\pm{0.24}$ & 2.92$\pm{0.09}$ & 0.51$\pm{0.01}$ & 8.48$\pm{0.55}$ & 0.48$\pm{0.01}$ & 0.37$\pm{0.02}$ & 78.98$\pm{0.85}$&3,7 \\ 
Pal 8* & -- & -- & -- & 6.22$\pm{0.86}$ & 1.97$\pm{0.92}$ & 0.52$\pm{0.17}$ & 1.36$\pm{0.37}$ & 0.11$\pm{0.04}$ & 0.78$\pm{0.17}$ & 25.92$\pm{2.62}$&-- \\ 
NGC 6681 & 2.68$\pm{0.81}$ & -3.78$\pm{0.91}$ & 220.3$\pm{0.9}$ & 6.77$\pm{0.94}$ & 0.39$\pm{0.08}$ & 0.89$\pm{0.03}$ & 5.16$\pm{0.62}$ & 0.13$\pm{0.02}$ & 0.25$\pm{0.03}$ & 22.22$\pm{1.56}$&8 \\ 
NGC 6712 & 4.2$\pm{0.4}$ & -2.0$\pm{0.4}$ & -107.6$\pm{0.5}$ & 6.4$\pm{0.27}$ & 0.83$\pm{0.06}$ & 0.77$\pm{0.02}$ & 4.49$\pm{0.24}$ & 0.27$\pm{0.02}$ & 0.51$\pm{0.03}$ & 41.35$\pm{1.13}$&3 \\ 
NGC 6715 & -1.75$\pm{0.43}$ & -4.95$\pm{0.6}$ & 141.3$\pm{0.3}$ & 160.26$\pm{52.92}$ & 15.64$\pm{0.76}$ & 0.82$\pm{0.05}$ & 29.87$\pm{1.73}$ & 0.55$\pm{0.01}$ & 0.02$\pm{0.01}$ & 287.11$\pm{13.8}$&8 \\ 
NGC 6717* & -- & -- & -- & 3.25$\pm{0.52}$ & 1.16$\pm{0.46}$ & 0.5$\pm{0.13}$ & 2.18$\pm{0.43}$ & 0.07$\pm{0.02}$ & 0.65$\pm{0.21}$ & 16.3$\pm{1.0}$&-- \\ 
NGC 6723 & -0.17$\pm{0.45}$ & -2.16$\pm{0.5}$ & -94.5$\pm{3.6}$ & 3.35$\pm{0.21}$ & 1.36$\pm{0.08}$ & 0.41$\pm{0.04}$ & 5.37$\pm{0.23}$ & 0.34$\pm{0.01}$ & 0.6$\pm{0.06}$ & 37.4$\pm{1.88}$&4 \\ 
NGC 6749 & 1.85$\pm{0.97}$ & -7.1$\pm{1.5}$ & -61.7$\pm{2.9}$ & 5.44$\pm{0.17}$ & 2.0$\pm{0.48}$ & 0.46$\pm{0.09}$ & 2.25$\pm{0.14}$ & 0.25$\pm{0.04}$ & 0.88$\pm{0.06}$ & 39.69$\pm{1.96}$&8 \\ 
NGC 6752 & -0.69$\pm{0.42}$ & -2.85$\pm{0.45}$ & -26.7$\pm{0.2}$ & 5.73$\pm{0.1}$ & 4.39$\pm{0.16}$ & 0.13$\pm{0.01}$ & 4.34$\pm{0.27}$ & 0.44$\pm{0.01}$ & 0.82$\pm{0.06}$ & 63.14$\pm{0.97}$&1,3 \\ 
NGC 6760 & 0.46$\pm{0.67}$ & -5.0$\pm{0.58}$ & -27.5$\pm{6.3}$ & 5.28$\pm{0.13}$ & 1.74$\pm{0.1}$ & 0.51$\pm{0.03}$ & 5.08$\pm{0.27}$ & 0.38$\pm{0.01}$ & 0.91$\pm{0.03}$ & 59.58$\pm{1.72}$&8 \\ 
NGC 6779 & 0.3$\pm{1.0}$ & 1.4$\pm{1.0}$ & -135.6$\pm{0.9}$ & 14.11$\pm{1.1}$ & 0.83$\pm{0.24}$ & 0.89$\pm{0.02}$ & 4.34$\pm{0.31}$ & 0.25$\pm{0.04}$ & 0.64$\pm{0.06}$ & 74.85$\pm{3.92}$&3 \\ 
Terzan 7* & -- & -- & -- & 19.14$\pm{3.2}$ & 8.39$\pm{2.7}$ & 0.42$\pm{0.12}$ & 0.44$\pm{0.04}$ & 0.29$\pm{0.05}$ & 0.62$\pm{0.28}$ & 50.3$\pm{3.6}$&-- \\ 
Pal 10* & -- & -- & -- & 7.31$\pm{0.54}$ & 2.26$\pm{0.94}$ & 0.54$\pm{0.15}$ & 1.34$\pm{0.28}$ & 0.16$\pm{0.05}$ & 0.85$\pm{0.11}$ & 34.33$\pm{2.04}$&-- \\ 
Arp 2 & -0.04$\pm{1.22}$ & -2.82$\pm{1.25}$ & 115.0$\pm{10.0}$ & 44.85$\pm{11.69}$ & 12.74$\pm{2.93}$ & 0.61$\pm{0.05}$ & 0.48$\pm{0.03}$ & 0.41$\pm{0.03}$ & 0.25$\pm{0.16}$ & 71.91$\pm{4.09}$&-- \\ 
NGC 6809 & -1.42$\pm{0.62}$ & -10.25$\pm{0.64}$ & 174.7$\pm{0.3}$ & 5.76$\pm{0.13}$ & 1.38$\pm{0.18}$ & 0.61$\pm{0.05}$ & 4.47$\pm{0.24}$ & 0.31$\pm{0.02}$ & 0.62$\pm{0.01}$ & 46.5$\pm{0.7}$&2,3 \\ 
Terzan 8* & -- & -- & -- & 24.12$\pm{4.09}$ & 10.67$\pm{3.55}$ & 0.41$\pm{0.12}$ & 0.43$\pm{0.04}$ & 0.33$\pm{0.05}$ & 0.58$\pm{0.28}$ & 60.69$\pm{4.08}$&8 \\ 
Pal 11 & -2.36$\pm{1.16}$ & 0.19$\pm{0.96}$ & -68.0$\pm{10.0}$ & 13.08$\pm{2.8}$ & 5.92$\pm{1.28}$ & 0.43$\pm{0.08}$ & 2.09$\pm{0.12}$ & 0.43$\pm{0.03}$ & 0.29$\pm{0.2}$ & 63.46$\pm{3.75}$&8 \\ 
NGC 6838 & -2.3$\pm{0.8}$ & -5.1$\pm{0.8}$ & -22.8$\pm{0.2}$ & 6.99$\pm{0.02}$ & 4.49$\pm{0.04}$ & 0.22$\pm{0.01}$ & 0.87$\pm{0.04}$ & 0.22$\pm{0.01}$ & 0.99$\pm{0.01}$ & 34.71$\pm{0.98}$&3 \\ 

\hline
\end{tabular}
\contcaption{}
\label{summary}
\end{table}
\end{landscape}

\begin{landscape}
\begin{table}
\footnotesize
\begin{tabular}{lccccccccccc}
\hline
Name & $\mu_{\alpha}$ & $\mu_{\delta}$ & V$_{\rm los}$ & R$_{apo}$ & R$_{peri}$ & ecc & M$_{i}$ & $\mu$ & $\phi$ & $r_J$ & Refs \\\\
 & mas/yr & mas/yr & km$\,$ s$^{-1}$ & kpc & kpc & & $10^5$ M$_{\odot}$ & & & pc & \\
\hline
\hline
NGC 6864 & -0.64$\pm{1.56}$ & -3.34$\pm{1.31}$ & -189.3$\pm{3.6}$ & 19.23$\pm{2.79}$ & 6.8$\pm{2.93}$ & 0.54$\pm{0.12}$ & 8.77$\pm{0.59}$ & 0.52$\pm{0.01}$ & 0.62$\pm{0.21}$ & 160.94$\pm{8.79}$&8 \\ 
NGC 6934 & 1.2$\pm{1.0}$ & -5.1$\pm{1.0}$ & -411.4$\pm{1.6}$ & 45.32$\pm{13.86}$ & 7.01$\pm{1.59}$ & 0.76$\pm{0.06}$ & 3.18$\pm{0.17}$ & 0.49$\pm{0.01}$ & 0.14$\pm{0.07}$ & 103.08$\pm{4.49}$&8 \\ 
NGC 6981 & -1.62$\pm{1.68}$ & -11.28$\pm{1.33}$ & -345.0$\pm{3.7}$ & 1841.6$\pm{243.73}$ & 10.66$\pm{0.46}$ & 0.99$\pm{0.01}$ & 2.12$\pm{0.09}$ & 0.5$\pm{0.01}$ & 0.0$\pm{0.01}$ & 90.82$\pm{2.96}$&8 \\ 
NGC 7006 & -0.96$\pm{0.35}$ & -1.14$\pm{0.4}$ & -384.1$\pm{0.4}$ & 118.52$\pm{35.85}$ & 18.14$\pm{4.14}$ & 0.75$\pm{0.02}$ & 3.8$\pm{0.25}$ & 0.52$\pm{0.01}$ & 0.2$\pm{0.1}$ & 233.22$\pm{10.63}$&8 \\ 
NGC 7078 & -0.95$\pm{0.51}$ & -5.63$\pm{0.5}$ & -107.0$\pm{0.2}$ & 11.08$\pm{0.43}$ & 6.3$\pm{0.81}$ & 0.28$\pm{0.04}$ & 15.29$\pm{0.96}$ & 0.53$\pm{0.01}$ & 0.89$\pm{0.07}$ & 158.56$\pm{5.95}$&3 \\ 
NGC 7089 & 5.9$\pm{0.86}$ & -4.95$\pm{0.86}$ & -5.3$\pm{2.0}$ & 41.84$\pm{10.67}$ & 5.84$\pm{0.82}$ & 0.76$\pm{0.05}$ & 13.16$\pm{0.78}$ & 0.53$\pm{0.01}$ & 0.12$\pm{0.05}$ & 149.59$\pm{5.55}$&3 \\ 
NGC 7099 & 1.42$\pm{0.69}$ & -7.71$\pm{0.65}$ & -184.2$\pm{0.2}$ & 7.45$\pm{0.21}$ & 2.82$\pm{0.42}$ & 0.45$\pm{0.06}$ & 3.56$\pm{0.18}$ & 0.38$\pm{0.02}$ & 0.95$\pm{0.03}$ & 67.86$\pm{2.72}$&2 \\ 
Pal 12 & -1.2$\pm{0.3}$ & -4.21$\pm{0.29}$ & 27.8$\pm{1.5}$ & 24.85$\pm{3.66}$ & 15.51$\pm{0.51}$ & 0.23$\pm{0.06}$ & 0.23$\pm{0.01}$ & 0.38$\pm{0.02}$ & 0.02$\pm{0.02}$ & 46.04$\pm{2.22}$&8 \\ 
Pal 13 & 2.3$\pm{0.26}$ & 0.27$\pm{0.25}$ & 25.2$\pm{0.3}$ & 153.46$\pm{34.77}$ & 12.35$\pm{1.23}$ & 0.85$\pm{0.02}$ & 0.14$\pm{0.01}$ & 0.28$\pm{0.03}$ & 0.1$\pm{0.03}$ & 49.69$\pm{2.46}$&8 \\ 
NGC 7492* & -- & -- & -- & 30.68$\pm{4.93}$ & 11.85$\pm{4.88}$ & 0.47$\pm{0.14}$ & 0.75$\pm{0.06}$ & 0.43$\pm{0.03}$ & 0.7$\pm{0.25}$ & 96.11$\pm{4.92}$&-- \\ 

\hline
\end{tabular}
\contcaption{}
\label{summary}
\end{table}
\end{landscape}

\bibliographystyle{mnras}

\bibliography{refs}

\label{lastpage}
\end{document}